\documentclass[11pt,a4paper]{article}
\usepackage{jheppub}
\usepackage{booktabs}
\usepackage{array}
\usepackage{mathtools}
\usepackage{placeins}
\usepackage{microtype}
\usepackage{orcidlink}
\let\manuscriptTableOfContents\tableofcontents
\renewcommand{\tableofcontents}{
 \begingroup\linespread{1}\small
 \manuscriptTableOfContents
 \endgroup
}

\hypersetup{
 breaklinks=true,
 hypertexnames=false,
 pdftitle={Complex spinodals and supercritical crossovers in Kerr-Newman-AdS4 black holes},
 pdfauthor={Bum-Hoon Lee, Hocheol Lee, and Somyadip Thakur},
 pdfkeywords={Kerr-Newman-AdS black holes, black hole chemistry, complex thermodynamics, Widom line, critical phenomena}
}

\newcommand{\dd}{\mathrm{d}}
\renewcommand{\Re}{\operatorname{Re}}
\renewcommand{\Im}{\operatorname{Im}}
\newcommand{\cG}{\mathcal G}
\newcommand{\cF}{\mathcal F}
\newcommand{\eps}{\epsilon}

\title{\boldmath Complex spinodals and supercritical crossovers\\
in Kerr--Newman--AdS$_4$ black holes}

\author[a,b,c]{Bum-Hoon Lee\,\orcidlink{0009-0008-3322-2087},}
\author[d]{Hocheol Lee\,\orcidlink{0000-0002-4420-5898},}
\author[a]{and Somyadip Thakur\,\orcidlink{0000-0003-2376-5906}}

\affiliation[a]{Center for Quantum Spacetime, Sogang University,
Seoul 04107, Republic of Korea}
\affiliation[b]{Department of Physics, Sogang University,
Seoul 04107, Republic of Korea}
\affiliation[c]{Department of Physics, Shanghai University,
99 Shangda Road, Baoshan District, Shanghai 200444, China}
\affiliation[d]{Department of Physics, Dongguk University,
Seoul 04620, Republic of Korea}

\emailAdd{bhl@sogang.ac.kr}
\emailAdd{insaying@dongguk.edu}
\emailAdd{smydp3thkr@gmail.com}

\abstract{
We study supercritical crossovers in four-dimensional Kerr--Newman--AdS
black holes at fixed angular momentum and charge.  From the exact
enthalpy, we prove uniqueness of the physical critical point and positivity
of the quartic Landau coefficient throughout the charge--rotation family,
treating the uncharged Kerr--AdS endpoint in separate scaling variables.
An exact quartic reduction determines the coexistence entropies and latent
heat throughout the subcritical regime.  Above criticality, the two
physical spinodals continue as a complex conjugate pair.  Their real
temperature projection defines a Widom-type line, and varying the
charge--rotation ratio assembles these lines into a crossover surface.
The Landau cusp fixes the imaginary entropy and temperature to scale with
the square root and three-halves powers of the pressure displacement from
criticality.  The projected fold and the local heat capacity ridge share
a critical tangent.  We derive their leading quadratic temperature
separation and prove that the ridge lies at a higher temperature near
criticality.  Exact endpoint solutions and numerical continuation check
these predictions.  The cusp exponents are common to the family, while the crossover temperature depends on the thermodynamic criterion used to define it.
}
\begin{document}
\maketitle
\flushbottom

\section{Introduction}
\label{sec:introduction}

When a first-order coexistence line terminates at a critical point, the two phases become continuously connected in the supercritical region. Thermodynamic response functions can nevertheless retain pronounced maxima, defining crossover curves beyond the endpoint of coexistence. Their locations depend on the criterion used, raising the question of how different crossover curves are related. Analytic continuation offers a complementary perspective: the real spinodals merge at criticality and continue into the complex plane. Understanding the relation between these complex folds and real thermodynamic response maxima is the central question of this work.

Charged anti-de Sitter (AdS) black holes exhibit van der Waals-type
criticality~\cite{Chamblin:1999,KubiznakMann:2012}.
In the extended thermodynamics, $P=-\Lambda/(8\pi)$ is the pressure and the mass is the enthalpy~\cite{CveticEtAl:2011,Dolan:2011,KubiznakMannTeo:2017}.
The canonical Kerr--Newman--AdS phase structure at fixed angular momentum
$J$ and charge $Q$ is well
established~\cite{CaldarelliEtAl:2000,ChengWeiLiu:2016}.
For $Q>0$, its reduced equation of state and coexistence curve depend on
\begin{equation}
 \eps=\frac{J}{Q^2},
 \label{eq:epsilon-intro}
\end{equation}
{\it{interpolating}} between Reissner--Nordstr\"om--AdS (RN--AdS) at $\eps=0$
and Kerr--AdS as $\eps\to\infty$~\cite{ChengWeiLiu:2016}.
The uncharged endpoint requires scaling with $J$ rather than $Q$.
We take $J\geq0$ without loss of generality.

Above the critical pressure, the real equation of state has a single
thermodynamic branch.  Using response function maxima to identify Widom
crossovers is established in fluid systems~\cite{XuEtAl:2005Widom}.
For AdS black holes, such curves have been studied through thermodynamic
curvature and heat capacity for rotating black holes~\cite{DehyadegariSheykhi:2021}
and through fluctuation measures for charged AdS black
holes~\cite{ZhaoEtAl:2025}.  Different diagnostics need not select the same
curve.

For Ising ferromagnets above their critical temperature, the partition function
zeros lie at imaginary magnetic fields.  In the thermodynamic limit, the
endpoints of the gap containing zero field define Yang--Lee edge
singularities of the analytically continued free energy.  Fisher described
the critical behavior at these edges using a scalar field theory with a
cubic interaction and an imaginary coupling~\cite{Fisher:1978Edge}.
At the mean-field level, the cubic form arises when two complex stationary
points merge: the first and second derivatives of the local potential
vanish, while the third remains nonzero.  This expansion about a complex
edge differs from the quartic Landau expansion about the real critical point.

Complex edge singularities have also been related to supercritical
crossovers in ordinary matter~\cite{OuyangYeLi:2024}.
For RN--AdS, a Widom line was defined by projecting complex singularities
onto the real temperature--pressure plane~\cite{XuMann:2026}.
Related constructions appear in charged Gauss--Bonnet and trace anomaly
gravity~\cite{LiChenWuXu:2025,OuZhangNie:2026}; the latter also compares
the projection with a fluctuation criterion and develops a Landau
expansion.  The study of complex saddles and Fisher zeros selected by the contour in Kerr--AdS$_5$ distinguishes local folds from the saddle
competition responsible for those zeros~\cite{Thakur:KerrAdS5Complex}.

Here the complex singularities are spinodals: folds of $S\mapsto T$ at
fixed $(P,J,Q)$, where the heat capacity diverges.  The pair merging at
criticality continues as complex conjugate roots.  Identifying these folds
with Lee--Yang or Fisher zeros would require a specified partition
function and integration contour, which we do not assume here\footnote{For related discussion we refer to \cite{Thakur:KerrAdS5Complex}}.

The local connection to real phase structure follows from Landau theory.
Volume based Landau descriptions of RN--AdS~\cite{XuWuYang:2021},
mean-field Kerr--AdS criticality~\cite{TsaiWuYang:2012}, and the off-shell
free energy $M-T_{\rm b}S$ at bath temperature
$T_{\rm b}$~\cite{YangEtAl:2022} are established.
We derive the entropy based coefficients for the full fixed-$(J,Q)$
family and use them to compare the complex projection with a real
heat capacity ridge.  Nonzero rotation gives a degree-eight fold equation,
so the relevant pair must be selected by continuation from the physical
real spinodals.

The quintic reduction proves that the physical critical point is unique
and that the quartic and pressure unfolding coefficients are strictly
positive for every finite $\eps\geq0$.  The independently scaled Kerr
endpoint is also nondegenerate.  An exact quartic reduction gives both
coexistence entropies and the latent heat
$L=T_{\rm coex}\Delta S=\Delta M$ without a slow rotation or critical
expansion. Finally, the projected fold and heat capacity ridge share a common tangent at the critical point but separate at quadratic order,
\begin{equation}
	\Delta_{\mathrm{ridge-fold}}
	=
	D_2(\epsilon)\,(p-1)^2
	+\mathcal{O}\!\left((p-1)^3\right),
	\qquad
	p=\frac{P}{P_c},
\end{equation}
where $\Delta_{\mathrm{ridge-fold}}=t_{C_P}-t_W$ and
$D_2(\epsilon)>0$.  The heat capacity ridge therefore lies at a higher
temperature near criticality, despite the common tangent.

The cusp fixes $|\Im z_*|\propto(p-1)^{1/2}$ and
$|\Im t_*|\propto(p-1)^{3/2}$, while their amplitudes depend on $\frac{J}{Q^2}$.
Numerical continuation extends the selected folds over the pressure
domain studied here.  Their real projections form one Widom-type line
for each ratio and a surface across the family; this construction
introduces no additional physical critical point at fixed $(J,Q)$.
Exact RN--AdS results and an independent Kerr--AdS calculation provide
endpoint checks.

\textbf{Organization of the paper}: Section~\ref{sec:thermodynamics} establishes the exact thermodynamics,
criticality, and coexistence.  Section~\ref{sec:landau} presents the Landau
normal form and its phase interpretation, and
Section~\ref{sec:complex-structure} develops the complex continuation and
response comparison.  Section~\ref{sec:discussion} discusses their
implications.  Appendix~\ref{app:rn} derives the RN--AdS coexistence
relations and compares the endpoints; Appendix~\ref{app:landau-derivation}
contains the detailed Landau derivation.

\section{Thermodynamics and real phase structure at fixed \texorpdfstring{$(J,Q)$}{(J,Q)}}
\label{sec:thermodynamics}

The exact enthalpy determines both the real phase structure and its
complex continuation.  We derive it from the horizon geometry, prove
uniqueness and nondegeneracy of the critical point, and obtain the
coexistence entropies and latent heat.

\subsection{Enthalpy, first law, and canonical free energy}
\label{sec:enthalpy-derivation}

We fix the geometric and thermodynamic conventions, then eliminate the
metric parameters in favor of $(S,J,Q,P)$ to derive the exact enthalpy
and the canonical potential.

\subsubsection{Einstein--Maxwell--AdS\texorpdfstring{$_4$}{4} geometry}

Four-dimensional Kerr--Newman--AdS is a stationary solution of
Einstein--Maxwell theory with negative cosmological constant
\cite{CaldarelliEtAl:2000}. Throughout this paper we use units $G=\hbar=c=k_{\rm B}=1$.
The Lorentzian bulk action is
\begin{equation}
I_{\rm bulk}
=
\frac{1}{16\pi}
\int \dd^4x\,\sqrt{-g}\,
\left(
R-2\Lambda-F_{\mu\nu}F^{\mu\nu}
\right),
\qquad
\Lambda=-\frac{3}{\ell^2},
\label{eq:EM-AdS-action}
\end{equation}
where $F_{\mu\nu}=\partial_\mu A_\nu-\partial_\nu A_\mu$.
The bulk action fixes the theory; a Euclidean action would additionally
require ensemble-dependent boundary terms and AdS counterterms\footnote{Although we derive the thermodynamic quantities from the Lorentzian geometry, they also govern the Euclidean saddle point description. At fixed \((P,J,Q)\), the renormalized Euclidean action, including the appropriate boundary terms and Legendre transformations, satisfies \(I_E^{(J,Q)}=\beta M-S\) on a regular black hole saddle, with \(\beta=T_H^{-1}\). Its leading semiclassical weight is therefore \(e^{S-\beta M}\), reproducing the equilibrium thermodynamics used here. For nonzero angular momentum, keeping the physical charges real generally requires a complex Euclidean continuation. This correspondence supports our thermodynamic treatment, but does not identify the local Landau potential with the full off-shell gravitational action or determine which complex saddles contribute to the partition function.}.

In extended black hole thermodynamics the cosmological constant is promoted
to a pressure,
\begin{equation}
P
=
-\frac{\Lambda}{8\pi}
=
\frac{3}{8\pi\ell^2},
\label{eq:pressure}
\end{equation}
whose conjugate variable is the thermodynamic volume.

In Boyer--Lindquist coordinates, the Kerr--Newman--AdS$_4$ line element
is~\cite{CaldarelliEtAl:2000}
\begin{align}
\dd s^2
={}&
-\frac{\Delta_r}{\rho^2}
\left(
\dd t-\frac{a\sin^2\theta}{\Xi}\,\dd\phi
\right)^2
+\frac{\rho^2}{\Delta_r}\,\dd r^2
+\frac{\rho^2}{\Delta_\theta}\,\dd\theta^2
\nonumber\\
&\quad
+\frac{\Delta_\theta\sin^2\theta}{\rho^2}
\left(
a\,\dd t-\frac{r^2+a^2}{\Xi}\,\dd\phi
\right)^2 ,
\label{eq:KNAdS-metric}
\end{align}
with
\begin{align}
\rho^2
&=
r^2+a^2\cos^2\theta,
&
\Delta_\theta
&=
1-\frac{a^2}{\ell^2}\cos^2\theta,
\nonumber\\
\Delta_r
&=
(r^2+a^2)
\left(
1+\frac{r^2}{\ell^2}
\right)
-2\mu r+q_0^2,
&
\Xi
&=
1-\frac{a^2}{\ell^2}.
\label{eq:KNAdS-functions}
\end{align}
We restrict to $\Xi>0$, or equivalently $a^2<\ell^2$.

For the purely electric solution, a gauge potential that vanishes at
infinity is
\begin{equation}
A
=
-\frac{q_0 r}{\rho^2}
\left(
\dd t-\frac{a\sin^2\theta}{\Xi}\,\dd\phi
\right).
\label{eq:gauge-potential}
\end{equation}
The parameter $q_0$ specifies the electric field; no magnetic charge is
included.

The event horizon lies at the largest positive root $r_+$ of
\begin{equation}
\Delta_r(r_+)=0,
\label{eq:horizon-condition}
\end{equation}
or explicitly,
\begin{equation}
(r_+^2+a^2)
\left(
1+\frac{r_+^2}{\ell^2}
\right)
-2\mu r_+
+q_0^2
=
0.
\label{eq:horizon-condition-expanded}
\end{equation}

The conserved charges, with the mass defined in a nonrotating frame at
the AdS boundary, are
\cite{CaldarelliEtAl:2000,ChengWeiLiu:2016}
\begin{equation}
M=\frac{\mu}{\Xi^2},
\qquad
J=\frac{a\mu}{\Xi^2}=aM,
\qquad
Q=\frac{q_0}{\Xi},
\label{eq:physical-charges}
\end{equation}
while the Bekenstein--Hawking entropy, with horizon area $A_+$, is
\begin{equation}
S
=
\frac{A_+}{4}
=
\frac{\pi(r_+^2+a^2)}{\Xi}.
\label{eq:entropy}
\end{equation}
\subsubsection{Derivation of the enthalpy}

To carry out this elimination explicitly, define
\begin{equation}
\sigma\equiv\frac{S}{\pi}.
\label{eq:sigma-definition}
\end{equation}
The entropy relation gives
\begin{equation}
r_+^2+a^2=\sigma\Xi,
\qquad
r_+^2
=
\sigma-a^2
\left(
1+\frac{\sigma}{\ell^2}
\right),
\label{eq:entropy-geometric-identities-a}
\end{equation}
and consequently
\begin{equation}
1+\frac{r_+^2}{\ell^2}
=
\Xi
\left(
1+\frac{\sigma}{\ell^2}
\right).
\label{eq:entropy-geometric-identities-b}
\end{equation}

Substituting
\begin{equation*}
\mu=M\Xi^2,
\qquad
q_0=Q\Xi
\end{equation*}
into the horizon equation and using
Eqs.~\eqref{eq:entropy-geometric-identities-a} and
\eqref{eq:entropy-geometric-identities-b}, the common factor of $\Xi^2$
cancels out and one obtains
\begin{equation}
2Mr_+
=
\sigma
\left(
1+\frac{\sigma}{\ell^2}
\right)
+Q^2.
\label{eq:physical-horizon-relation}
\end{equation}
Squaring this relation and eliminating the rotation parameter through
$a=J/M$ yields
\begin{equation}
4\sigma M^2
-
4J^2
\left(
1+\frac{\sigma}{\ell^2}
\right)
=
\left[
\sigma
\left(
1+\frac{\sigma}{\ell^2}
\right)
+Q^2
\right]^2,
\label{eq:enthalpy-elimination}
\end{equation}
or equivalently
\begin{equation}
M^2
=
\frac{
\left(
\sigma+Q^2+\sigma^2/\ell^2
\right)^2
+
4J^2
\left(
1+\sigma/\ell^2
\right)
}
{4\sigma}.
\label{eq:geometric-mass-squared}
\end{equation}

Using
\begin{equation*}
\sigma=\frac{S}{\pi},
\qquad
\ell^{-2}=\frac{8\pi P}{3},
\end{equation*}
and choosing the positive mass branch, we obtain the exact enthalpy
\begin{equation}
{
M(S,J,Q,P)
=
\frac{1}{2}
\sqrt{
\frac{
\left(
S+\pi Q^2+\frac{8}{3}PS^2
\right)^2
+
4\pi^2J^2
\left(
1+\frac{8}{3}PS
\right)
}
{\pi S}
}
}.
\label{eq:enthalpy}
\end{equation}
No expansion in $J$, $Q$, or $P$ has been made. The limits $J\to0$ and
$P\to0$ reproduce, respectively, the Reissner--Nordstr\"om--AdS enthalpy
and the asymptotically flat Kerr--Newman mass formula.

\subsubsection{First law and canonical thermodynamics}

Allowing the pressure to vary identifies the mass with the enthalpy,
\begin{equation*}
M=H=U+PV.
\end{equation*}
The extended first law and Smarr relation are
\cite{CveticEtAl:2011,Dolan:2011}
\begin{align}
\dd M
&=
T\,\dd S+\Omega\,\dd J+\Phi_Q\,\dd Q+V\,\dd P,
\label{eq:first-law}
\\
M
&=
2(TS+\Omega J-PV)+\Phi_Q Q .
\label{eq:smarr}
\end{align}
The equilibrium temperature $T=T_H$ and the other conjugate variables
are obtained directly from the enthalpy:
\begin{equation}
\begin{aligned}
T
&=
\left(
\frac{\partial M}{\partial S}
\right)_{J,Q,P},
&\qquad \Omega
=
\left(
\frac{\partial M}{\partial J}
\right)_{S,Q,P},\\
\Phi_Q
&=
\left(
\frac{\partial M}{\partial Q}
\right)_{S,J,P},
&\qquad V
=
\left(
\frac{\partial M}{\partial P}
\right)_{S,J,Q}.
\end{aligned}
\label{eq:conjugate-variables}
\end{equation}
When the enthalpy is differentiated with respect to $S$ at fixed physical angular
momentum $J$, the rotation parameter $a=J/M$ varies with the mass\footnote{The Smarr formula follows equivalently from the homogeneity relation
\begin{equation}
M(\alpha^2S,\alpha^2J,\alpha Q,\alpha^{-2}P)
=
\alpha\,M(S,J,Q,P),
\label{eq:homogeneity}
\end{equation}
by differentiating with respect to $\alpha$ at $\alpha=1$.}.

At fixed $(J,Q,P)$, the equilibrium thermodynamic potential is
\begin{equation}
\mathcal G(T;J,Q,P)
=
M-TS,
\qquad
\dd\mathcal G
=
-S\,\dd T+\Omega\,\dd J+\Phi_Q\,\dd Q+V\,\dd P.
\label{eq:gibbs}
\end{equation}
We call $\mathcal G$ the canonical Gibbs free energy.  Below criticality,
its swallowtail encodes coexistence through equal values on the stable
branches; turning points of $T(S)$ instead locate the spinodals.

\subsubsection{Local canonical function and stability}

For the local analysis, retain entropy as an independent variable and define
\begin{equation}
\Phi_{J,Q,P}(S;\beta)
=
S-\beta M(S,J,Q,P),
\qquad
\beta>0,
\label{eq:local-canonical-function}
\end{equation}
where the inverse bath temperature $\beta=1/T_{\rm b}$ is independent of
entropy.  The corresponding off-shell Gibbs potential is
\begin{equation}
\begin{aligned}
\mathcal G_{\rm off}(S;T_{\rm b},J,Q,P)
&=
M(S,J,Q,P)-T_{\rm b}S,
\\
\Phi_{J,Q,P}
&=
-\beta\,\mathcal G_{\rm off},
\qquad
T_{\rm b}=\beta^{-1}.
\end{aligned}
\label{eq:off-shell-relation}
\end{equation}

Its stationary equation is
\begin{equation}
\frac{\partial\Phi_{J,Q,P}}{\partial S}
=
1-\beta
\left(
\frac{\partial M}{\partial S}
\right)_{J,Q,P}
=
1-\beta T_H(S,J,Q,P)
=
0,
\label{eq:local-stationarity}
\end{equation}
and gives $T_H=T_{\rm b}$ at equilibrium.

At a stationary point,
\begin{align}
\frac{\partial^2\Phi_{J,Q,P}}{\partial S^2}
&=
-\beta
\left(
\frac{\partial T_H}{\partial S}
\right)_{J,Q,P}
=
-\frac{1}{C_{P,J,Q}},
\label{eq:hessian-heat-capacity}
\\
C_{P,J,Q}
&=
T
\left(
\frac{\partial S}{\partial T}
\right)_{P,J,Q}.
\end{align}
Thus a nondegenerate local maximum of $\Phi_{J,Q,P}$ corresponds to positive
canonical heat capacity. The spinodal condition is
\begin{equation}
\left(
\frac{\partial T_H}{\partial S}
\right)_{J,Q,P}
=
0,
\label{eq:spinodal-preliminary}
\end{equation}
for which the quadratic curvature of the local potential vanishes. At the
critical point the degeneracy is stronger, and higher derivatives determine
the local normal form.

This tests only entropy fluctuations at fixed $(J,Q,P)$, not charge or
dynamical stability.  We assign no integration contour or partition
function to $\Phi_{J,Q,P}$; its local expansion gives the Landau potential
in Section~\ref{sec:cusp}.

\subsection{Reduced thermodynamics and spinodal folds}
\label{sec:reduced-thermodynamics}

We express the thermodynamics in dimensionless variables and identify the
polynomial that controls the heat capacity and spinodal folds.

\subsubsection{Dimensionless charge--rotation reduction}
\label{sec:reduction}

Scaling out the charge isolates the parameter controlling the
interpolation between charged and rotating black holes.  This reduction
also makes the physical domain and the possible complex branch loci
explicit.

Assume $Q>0$ and take $J\geq0$ without loss of generality.  Define
\begin{equation}
 \begin{aligned}
  s&=\frac{S}{\pi Q^2},
  &\qquad \lambda&=8\pi P Q^2,\\
  \eps&=\frac{J}{Q^2},
  &\qquad \Theta&=4\pi Q T,
 \end{aligned}
 \label{eq:reduced-variables}
\end{equation}
and
\begin{equation}
 \mathfrak m=\frac{M}{Q},
 \qquad
 \mathfrak g=\frac{\cG}{Q},
 \qquad
 \mathfrak v=\frac{3V}{4\pi Q^3}.
 \label{eq:reduced-potentials}
\end{equation}
The reduced functions can be written in terms of two polynomials:
\begin{align}
 \mathcal D(s,\lambda;\eps)
 &=12\eps^2(\lambda s+3)
 +(\lambda s^2+3s+3)^2,
 \label{eq:D-def}\\
 \mathcal N(s,\lambda;\eps)
 &=\lambda^2s^4+4\lambda s^3+(2\lambda+3)s^2
 -12\eps^2-3.
 \label{eq:N-def}
\end{align}
The enthalpy, temperature, Gibbs free energy, and thermodynamic volume become
\begin{align}
 \mathfrak m
 &=\frac{\sqrt{\mathcal D}}{6\sqrt{s}},
 \label{eq:reduced-mass}\\
 \Theta
 &=\frac{\mathcal N}{s^{3/2}\sqrt{\mathcal D}},
 \label{eq:reduced-temperature}\\
 \mathfrak g
 &=\frac{\sqrt{\mathcal D}}{6\sqrt{s}}-\frac{s\Theta}{4},
 \label{eq:reduced-gibbs}\\
 \mathfrak v
 &=\frac{\sqrt{s}\left[6\eps^2+s(\lambda s^2+3s+3)\right]}
 {\sqrt{\mathcal D}}.
 \label{eq:reduced-volume}
\end{align}
All reduced thermodynamic state functions therefore depend on $Q$ and $J$
only through $\eps$, in agreement with Ref.~\cite{ChengWeiLiu:2016}.

On the real slice, $s>0$ and $\lambda>0$ imply $\mathcal D>0$.  The physical
nonextremal domain additionally has $\Theta>0$, equivalently
$\mathcal N>0$.  Its lower endpoint is the positive solution of
$\mathcal N=0$.  Upon complexification, $s=0$ and, generically, the simple zeros of
$\mathcal D$ are intrinsic branch loci of these expressions.  They must be
excluded when interpreting roots of a fold equation obtained by clearing the denominators.
At $\eps=0$, $\mathcal D$ is a perfect square and its apparent branch
loci cancel in the temperature; Appendix~\ref{app:rn} gives the reduced
fold equation explicitly.

\subsubsection{Heat capacity and the eighth-degree fold equation}
\label{sec:fold-polynomial}

Heat capacity divergences and spinodal folds both occur at zeros of
$\Theta_s$.  Clearing its denominator gives the polynomial equation used
for complex continuation.

Differentiating Eq.~\eqref{eq:reduced-temperature} gives
\begin{equation}
 \Theta_s
 =\frac{\cF(s,\lambda;\eps)}
 {2s^{5/2}\mathcal D^{3/2}}.
 \label{eq:theta-derivative}
\end{equation}
Here variable subscripts denote partial derivatives, for example
$\Theta_s=(\partial\Theta/\partial s)_{\lambda,\eps}$ and
$\Theta_{s\lambda}=\partial^2\Theta/(\partial s\,\partial\lambda)$
at fixed $\eps$. The label $c$ instead denotes evaluation at the critical
point. \begin{equation}
 {
 \cF
 =2s\mathcal D\,\mathcal N_s
 -\mathcal N\left(3\mathcal D+s\mathcal D_s\right).
 }
 \label{eq:F-def}
\end{equation}
The heat capacity is consequently
\begin{equation}
 {
 \frac{C_{P,J,Q}}{\pi Q^2}
 =\frac{2s\mathcal D\mathcal N}{\cF}.
 }
 \label{eq:heat-capacity}
\end{equation}
For fixed $\lambda>0$ and real $\eps$, $\cF$ is a polynomial of degree
eight in $s$.  Substituting Eqs.~\eqref{eq:D-def} and~\eqref{eq:N-def}
into Eq.~\eqref{eq:F-def} gives the canonical fold equation explicitly:
\begin{equation}
 \begin{aligned}
 \cF(s,\lambda;\eps)={}&
 \lambda^4s^8+8\lambda^3s^7+6\lambda^2(3+2\lambda)s^6\\
 &+24\lambda^2(3+2\eps^2\lambda)s^5
 +9\bigl[(6+40\eps^2)\lambda^2+12\lambda-3\bigr]s^4\\
 &+216\lambda(1+4\eps^2)s^3
 +54(1+4\eps^2)(3+2\lambda)s^2\\
 &+72(1+4\eps^2)(3+2\eps^2\lambda)s
 +81(1+4\eps^2)^2=0.
 \end{aligned}
 \label{eq:fold-equation}
\end{equation}
Only roots away from the intrinsic branch loci are admissible.
The critical point is the double root
selected by
\begin{equation}
 \cF(s_c,\lambda_c;\eps)=0,
 \qquad
 \cF_s(s_c,\lambda_c;\eps)=0.
 \label{eq:critical-equations}
\end{equation}
Equivalently, $\Theta_s=\Theta_{ss}=0$.  The polynomial has eight roots counted with multiplicity.  At the
RN endpoint, six arise from a factor that cancels from $\Theta_s$; for
finite rotation all candidates must still be checked against the intrinsic
branch loci.  The relevant pair consists of the two positive real roots
connected to the physical subcritical spinodals.
This continuation criterion will be essential above criticality.

\subsection{Physical critical point and endpoint limits}
\label{sec:critical-point}

We establish uniqueness and nondegeneracy of the critical point, then
specify its scales across the charge--rotation family and at both endpoints.

\subsubsection{Uniqueness and nondegeneracy of the physical critical point}
\label{sec:analytic-critical-proof}

The squared reduced enthalpy reduces the criticality conditions to a
single polynomial and determines the signs of the derivatives required
for a generic cusp.

Define
\begin{align}
 h(s,\lambda;\eps)&=\mathfrak m^2
 =\frac{\lambda^2s^3}{36}+\frac{\lambda s^2}{6}
  +\frac{3+2\lambda}{12}s+\frac12+\frac{\lambda\eps^2}{3}
  +\frac{1+4\eps^2}{4s},\nonumber\\
 \mathcal A&=2h h_{ss}-h_s^2.
 \label{eq:squared-enthalpy}
\end{align}
For real $s,\lambda>0$, $h>0$ and
\begin{equation}
 \Theta=\frac{2h_s}{\sqrt h},\qquad
 \Theta_s=\frac{\mathcal A}{h^{3/2}},\qquad
 \mathcal A_s=2h h_{sss},\qquad
 \cF=432s^4\mathcal A.
\end{equation}
Consequently a critical point obeys $h_{sss}=0$.  With
$r=\sqrt{1+4\eps^2}\geq1$, this fixes
\begin{equation}
 \lambda_c s_c^2=3r.
 \label{eq:analytic-critical-relation}
\end{equation}
Substitution into $\mathcal A=0$ gives the single quintic equation
\begin{align}
 f_r(s_c)&=0,\nonumber\\
 f_r(s)&=s^5-12r(r+1)s^3-32r^2(r+1)s^2\nonumber\\
 &\quad-12r^2(r+1)(3r-1)s-16r^3(r^2-1).
 \label{eq:critical-quintic}
\end{align}
Its coefficients have exactly one sign change, so there is precisely one
positive root for every $r\geq1$.  At $r=1$,
$f_1(s)=s(s-6)(s+2)^3$, recovering the RN critical point.
At the positive root $f'_r(s_c)>0$: use $f_r(s_c)=0$ to replace the leading
term in $s_c f'_r(s_c)$ by a sum of strictly positive terms of lower degree.

This result also fixes the number of real spinodals. For fixed
$\lambda>0$, $\mathcal A_s$ changes sign only at
$s_0=\sqrt{3r/\lambda}$, from negative to positive, and
$\mathcal A\to+\infty$ as either $s\to0^+$ or $s\to\infty$.
Along $\lambda=3r/s^2$ one has
\begin{equation}
 \mathcal A(s,3r/s^2;\eps)=-\frac{f_r(s)}{16s^5}.
\end{equation}
Thus its minimum is negative, zero, or positive according as
$\lambda<\lambda_c$, $\lambda=\lambda_c$, or $\lambda>\lambda_c$.
There are exactly two positive real folds below criticality, one double
fold at criticality, and none above.  Both subcritical folds are
nonextremal.  Indeed $h_{ss}>0$, the unique extremal point has $h_s=0$
and $\mathcal A=2hh_{ss}>0$, whereas
$h_s(s_0)=1/4+r/s_0+r(r+1)/(2s_0^2)>0$.

The same identities prove that the critical point is always a generic
cusp.  At criticality,
\begin{equation}
 \Theta_{sss}\big|_c
 =\frac{2h_{ssss,c}}{\sqrt{h_c}}
 =\frac{12r^2}{s_c^5\sqrt{h_c}}>0,
 \qquad
 \Theta_{s\lambda}\big|_c
 =\frac{f'_r(s_c)}{32\lambda_c s_c^4h_c^{3/2}}>0.
 \label{eq:analytic-cusp-positivity}
\end{equation}
The second identity follows by differentiating
$\mathcal A(s,3r/s^2)=-f_r(s)/(16s^5)$ at its zero and using
$\mathcal A_s|_c=0$.  Hence the quartic coefficient in the local
off-shell potential and the pressure unfolding coefficient are strictly
positive for every finite $\eps\geq0$.

\subsubsection{Critical scales and the RN--AdS and Kerr--AdS endpoints}
\label{sec:critical-family}

Critical normalization allows comparison of coexistence curves and the
Landau coefficients in Section~\ref{sec:landau} across the family.  The
RN--AdS and Kerr--AdS endpoints require charge and angular momentum
scales, respectively.

At $\eps=0$, Eqs.~\eqref{eq:critical-equations} give the exact RN--AdS
values
\begin{equation}
 s_c=6,
 \qquad
 \lambda_c=\frac{1}{12},
 \qquad
 \Theta_c=\frac{2\sqrt6}{9}.
 \label{eq:rn-critical}
\end{equation}
The opposite endpoint requires a change of scale because $Q\to0$ at fixed
$J$ sends $\eps\to\infty$.  Define
\begin{equation}
 u=\frac{s}{\eps}=\frac{S}{\pi J},
 \qquad
 \bar\lambda=\eps\lambda=8\pi PJ,
 \qquad
 \bar\Theta=\sqrt\eps\,\Theta=4\pi\sqrt J\,T.
 \label{eq:kerr-rescaling}
\end{equation}
The polynomials in the Kerr limit are
\begin{align}
 \mathcal D_K
 &=12(\bar\lambda u+3)+u^2(\bar\lambda u+3)^2,
 \label{eq:D-kerr}\\
 \mathcal N_K
 &=\bar\lambda^2u^4+4\bar\lambda u^3+3u^2-12,
 \label{eq:N-kerr}
\end{align}
and $\bar\Theta=\mathcal N_K/(u^{3/2}\sqrt{\mathcal D_K})$.  Solving the
critical equations gives
\begin{equation}
 u_c=9.141501127,
 \qquad
 \bar\lambda_c=0.07179863968,
 \qquad
 \bar\Theta_c=0.5246337238,
 \label{eq:kerr-critical}
\end{equation}
consistent with the exact four-dimensional Kerr--AdS critical
scales~\cite{WeiChengLiu:2016}.
In the Kerr variables, the same argument gives
\begin{align}
 \bar\lambda_c u_c^2&=6,\nonumber\\
 u_c^5-48u_c^3-256u_c^2-576u_c-512&=0.
 \label{eq:kerr-critical-quintic}
\end{align}
This polynomial again has exactly one positive root.  The squared Kerr
enthalpy is
$h_K=\bar\lambda^2u^3/36+\bar\lambda u^2/6+u/4+
\bar\lambda/3+1/u$, so the positivity proof applies directly to this
endpoint as well.

For convenience of  plotting the complete family on finite axes, we use
\begin{equation}
 \chi=\frac{\eps}{1+\eps},
 \qquad
 \widehat s_c=\frac{s_c}{1+\eps},
 \qquad
 \widehat\lambda_c=(1+\eps)\lambda_c,
 \qquad
 \widehat\Theta_c=\sqrt{1+\eps}\,\Theta_c.
 \label{eq:compact-critical-variables}
\end{equation}
This compactification is a visualization convention, not a new
thermodynamic scaling.

For $\eps=1$, restoring $Q$ gives
\begin{equation}
 P_c=0.00191841494\,Q^{-2},
 \qquad
 S_c=37.05627809\,Q^2,
 \qquad
 T_c=0.03350968616\,Q^{-1}.
 \label{eq:epsilon-one-critical}
\end{equation}

Critical data for representative values of $\eps$ are collected with
the Landau amplitudes in Table~\ref{tab:critical-benchmarks}.

\subsection{Coexistence and latent heat}
\label{sec:coexistence-latent-heat}

We identify the stable branches and derive the coexistence entropies and
latent heat from equality of their Gibbs free energies.

\subsubsection{Branches, coexistence, and local stability}
\label{sec:real-branches}

The sign of the heat capacity identifies locally stable branches, and
equal Gibbs free energies determine coexistence.  The bounding spinodals
are continued into the complex plane in Section~\ref{sec:complex-structure}.

At fixed $\eps$ introduce critical normalization
\begin{equation}
 p=\frac{P}{P_c}=\frac{\lambda}{\lambda_c},
 \qquad
 t=\frac{T}{T_c}=\frac{\Theta}{\Theta_c},
 \qquad
 z=\sqrt{\frac{s}{s_c}}.
 \label{eq:critical-normalization}
\end{equation}
Here $s/s_c=S/S_c$, so $z$ is the square root of the reduced entropy.
For comparisons with entropy based phase diagrams we use
$\widetilde S=S/S_c=z^2$ and
$\widetilde G=\cG/\cG_c$, where
$\cG_c=\cG(S_c,P_c,J,Q)$ is held fixed when comparing isobars.
The Gibbs curves are traced parametrically in entropy, retaining all
branches of $(t(S),\widetilde G(S))$.
For $p<1$, the map $s\mapsto\Theta$ has two turning points.  Since
$\mathcal N>0$ in the nonextremal region, Eq.~\eqref{eq:heat-capacity}
shows that the intermediate branch has negative heat capacity while the
outer branches are locally stable.  The coexistence temperature is obtained
from two entropies $s_< <s_>$ satisfying
\begin{equation}
 \Theta(s_<,\lambda;\eps)=\Theta(s_>,\lambda;\eps),
 \qquad
 \mathfrak g(s_<,\lambda;\eps)=\mathfrak g(s_>,\lambda;\eps).
 \label{eq:coexistence-equations}
\end{equation}

\begin{figure}[tbp]
 \centering
 \includegraphics[width=\textwidth]{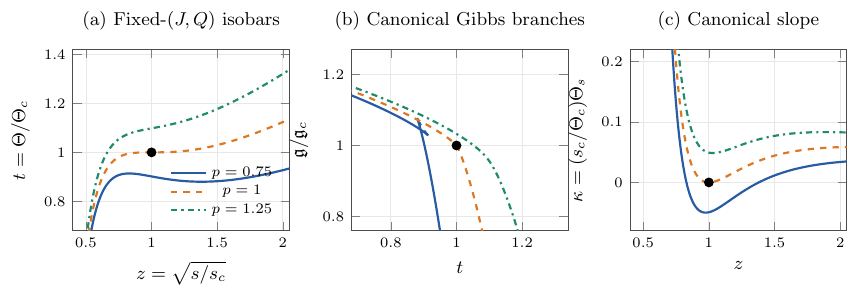}
 \caption{Real fixed-$(J,Q)$ thermodynamics at $\eps=1$ and
 $p=0.75,1,1.25$ in all three panels.
 (a) Reduced isobars plotted against $z=\sqrt{S/S_c}$.
 (b) Parametric canonical Gibbs branches, normalized by the fixed critical
 value $\cG_c$; the dashed part of the subcritical curve is locally unstable.
 (c) The normalized canonical slope
 $\kappa=(s_c/\Theta_c)\Theta_s$.  It has two simple zeros for $p<1$, a
 double zero at $p=1$, and no real zero for $p>1$.}
 \label{fig:real-thermodynamics}
\end{figure}

Figure~\ref{fig:real-thermodynamics} displays these branches at $\eps=1$.
The two spinodals merge at $p=1$, above which the real equation of state
is monotone.  Figure~\ref{fig:endpoint-benchmark} checks the equation of
state, Gibbs normalization, and branch selection against the established
RN--AdS and Kerr--AdS phase structures~\cite{ChengWeiLiu:2016}.

\subsubsection{Latent heat for Kerr--Newman--AdS}
\label{sec:latent-heat}

The coexistence conditions in Eq.~\eqref{eq:coexistence-equations}
can be reduced algebraically to determine both entropies and the latent
heat throughout the subcritical Kerr--Newman--AdS family.  We derive this
exact result from the common tangent to the enthalpy, retaining both
angular momentum and charge.  Its near-critical limit will follow from
the Landau potential in Section~\ref{sec:landau-phase-structure}.

At fixed $(P,J,Q)$, equality of the Gibbs free energies of the small and
large black holes implies
\begin{equation}
 0=\cG_>-\cG_<
  =M_>-M_<-T_{\rm coex}(S_>-S_<).
 \label{eq:latent-equal-gibbs}
\end{equation}
The latent heat is therefore
\begin{equation}
 {L=T_{\rm coex}(S_>-S_<)=M_>-M_<.}
 \label{eq:latent-exact}
\end{equation}
The entropies here are the coexistence values, not the spinodals.
Since $M$ is the enthalpy, its jump includes the pressure--volume work.

An exact reduction follows from the common tangent to $M(S)$ at
coexistence.  Write $S_<,S_>$ for its two contact points and
$\cG_{\rm coex}$ for their common Gibbs energy.  Both $M$ and its entropy
derivative agree with the line $\cG_{\rm coex}+T_{\rm coex}S$ at these
points.  Since $S M^2$ is a quartic polynomial, the two double zeros give
\begin{equation}
 S\bigl[M(S,J,Q,P)^2-(\cG_{\rm coex}+T_{\rm coex}S)^2\bigr]
 =\frac{16P^2}{9\pi}(S-S_<)^2(S-S_>)^2.
 \label{eq:latent-common-tangent}
\end{equation}
By matching coefficients, and defining
\begin{equation}
 R=\sqrt{Q^4+4J^2},\qquad k=8P(S_<+S_>),\qquad
 S_<S_>=\frac{3\pi R}{8P},
 \label{eq:latent-sum-product}
\end{equation}
yields a single quartic equation for $k$:
\begin{equation}
 \bigl[9-k^2+48\pi P(Q^2-R)\bigr]^2
 =384\pi P(k+3)\bigl(3Q^2+16\pi PJ^2+Rk\bigr).
 \label{eq:latent-full-quartic}
\end{equation}
For $0<P<P_c$, the physical root satisfies
\begin{equation}
 k>\sqrt{96\pi PR},\qquad
 9-k^2+48\pi P(Q^2-R)>0.
 \label{eq:latent-root-selection}
\end{equation}
The second condition restores the positive sign of the unsquared
coefficient relation.  On this domain its left-hand side decreases with
$k$, while the positive square root of the right-hand side of
Eq.~\eqref{eq:latent-full-quartic} increases, so the admissible root is
unique.  The resulting coexistence data are
\begin{align}
 S_{\lessgtr}&=\frac{k\mp\sqrt{k^2-96\pi PR}}{16P},
 & T_{\rm coex}&=\frac23\sqrt{\frac{P(k+3)}{\pi}},
 \label{eq:latent-full-states}\\
 \cG_{\rm coex}&=\sqrt{\frac{3Q^2+16\pi PJ^2+Rk}{6}},
 & L&=\frac1{12}\sqrt{\frac{(k+3)(k^2-96\pi PR)}{\pi P}}.
 \label{eq:latent-full-result}
\end{align}
These formulas retain the full dependence on $J,Q,P$ and apply along the
small/large coexistence curve, without a critical or slow rotation
expansion.  They also remain regular at $Q=0$ with $J\ne0$.
The case $J=Q=0$ has no such small/large coexistence curve and is excluded.
Solving Eq.~\eqref{eq:latent-full-quartic} subject to
Eq.~\eqref{eq:latent-root-selection} gives, for example, at $J=Q=1$
and $P=10^{-3}<P_c$,
\begin{equation}
 \begin{aligned}
 S_<&=13.79164205, & S_>&=191.0073880,\\
 T_{\rm coex}&=0.02561636722, & L&=4.539623625.
 \end{aligned}
 \label{eq:latent-full-example}
\end{equation}

To compare the charge--rotation family, define
\begin{equation}
 \mathcal L(p;\eps)=\frac{L}{T_cS_c}
 =t_{\rm coex}\frac{s_>-s_<}{s_c}
 =\frac{4\,[\mathfrak m(s_>,p\lambda_c;\eps)
                  -\mathfrak m(s_<,p\lambda_c;\eps)]}{\Theta_c s_c}.
 \label{eq:latent-reduced}
\end{equation}
We evaluate Eq.~\eqref{eq:latent-full-result} and independently solve
Eq.~\eqref{eq:coexistence-equations} on the two stable branches.
Agreement of the two methods and of the entropy and enthalpy expressions
in Eq.~\eqref{eq:latent-reduced} checks the algebraic reduction.  Table~\ref{tab:latent-heat} gives representative
values.  At $Q=0$, the same dimensionless definition applies, with
$s/s_c$ replaced by $u/u_c$ and the mass and temperature scaled by $\sqrt J$
as in Section~\ref{sec:critical-family}.

\begin{table}[tbp]
 \centering
 \begin{tabular}{crrrr}
 \toprule
 $p$ & RN--AdS ($\eps=0$) & $\eps=1$ & $\eps=10$ & Kerr--AdS\\
 \midrule
 0.50 & 4.298621 & 3.851588 & 3.463201 & 3.408507\\
 0.80 & 1.873600 & 1.708933 & 1.562416 & 1.541499\\
 0.95 & 0.809924 & 0.745608 & 0.687348 & 0.678951\\
 0.99 & 0.349466 & 0.322526 & 0.297982 & 0.294434\\
 \bottomrule
 \end{tabular}
 \caption{Latent heat of exact coexistence, $\mathcal L=L/(T_cS_c)$, rounded
 to six decimal places.  Each column uses its own critical scales.  The
 Kerr column is computed directly at $Q=0$, rather than approximated by
 a large finite $\eps$.}
 \label{tab:latent-heat}
\end{table}

For RN--AdS, $R=Q^2$ and Eq.~\eqref{eq:latent-full-quartic} reduces to
$(k+3)^2[(3-k)^2-384\pi PQ^2]=0$.  Its physical root is
$k=3-2\sqrt p$, which reproduces the coexistence relation in
Appendix~\ref{app:rn} and gives
\begin{equation}
 t_{\rm coex}^{\rm RN}=\sqrt{\frac{p(3-\sqrt p)}{2}},\qquad
 \mathcal L_{\rm RN}=(3-\sqrt p)
       \sqrt{\frac{6(1-\sqrt p)}{p}},\qquad 0<p<1.
 \label{eq:latent-rn}
\end{equation}

The coexistence curves in Fig.~\ref{fig:phase-structure}(a) reproduce
the interpolation of Ref.~\cite{ChengWeiLiu:2016}; Appendix~\ref{app:rn}
gives the endpoint checks.

\FloatBarrier

\section{Landau theory of the critical point}
\label{sec:landau}

The exact enthalpy determines a local quartic potential whose stationary
points describe coexistence, metastability, and the continuation of the
spinodals through criticality.  We present its normal form and physical
consequences here; Appendix~\ref{app:landau-derivation} derives the Taylor
expansion, normalization, and stationary solutions.

\subsection{Off-shell Landau potential and the canonical cusp}
\label{sec:cusp}

Keeping the bath temperature independent of entropy gives a potential
whose extrema are the equilibrium black hole branches.  The critical
conditions then fix its leading entropy dependence to be quartic.

For $x=s-s_c$, $\delta\lambda=\lambda_c\delta p$, and $\delta p=p-1$,
introduce $t_{\rm b}=T_{\rm b}/T_c$ and the off-shell potential
\cite{YangEtAl:2022}
\begin{equation}
 {
 \mathfrak g_{\rm off}(s;p,t_{\rm b},\eps)
 =\mathfrak m(s,p\lambda_c;\eps)
 -\frac{\Theta_c t_{\rm b}}{4}s .
 }
 \label{eq:off-shell-gibbs}
\end{equation}
At fixed $(p,t_{\rm b},\eps)$, the first law gives
\begin{equation}
 4\,\partial_s\mathfrak g_{\rm off}
 =\Theta-\Theta_c t_{\rm b},
 \qquad
 4\,\partial_s^2\mathfrak g_{\rm off}=\Theta_s.
 \label{eq:off-shell-stationarity}
\end{equation}
In the physical nonextremal domain with $T_{\rm b}>0$, a nondegenerate
minimum has $C_{P,J,Q}>0$ and a maximum has $C_{P,J,Q}<0$; the critical
minimum is quartic with divergent heat capacity.  The relation
\begin{equation}
 \Phi_{J,Q,P}(S;\beta)=-\beta Q\,\mathfrak g_{\rm off},
 \qquad \beta^{-1}=T_{\rm b},
 \label{eq:phi-goff-relation}
\end{equation}
reverses the curvature signs without changing the stationary branches.
Substituting $T_{\rm b}=T(s)$ before differentiation instead gives the
on-shell Gibbs function, whose entropy derivative is $-s\Theta_s/4$.

The cusp scaling is
\begin{equation}
 x=O(\rho),\qquad \delta p=O(\rho^2),\qquad
 \delta t_{\rm b}-B\delta p=O(\rho^3),
 \qquad \delta t_{\rm b}=t_{\rm b}-1.
 \label{eq:cusp-power-counting}
\end{equation}
With derivatives evaluated at criticality, the temperature and fold
condition become
\begin{align}
 \Theta-\Theta_c
 &=\Theta_\lambda\delta\lambda
 +\Theta_{s\lambda}x\delta\lambda
 +\frac{\Theta_{sss}}6x^3+O_{\rm w}(\rho^4),
 \label{eq:cusp-expansion}\\
 0=\Theta_s
 &=\Theta_{s\lambda}\delta\lambda
 +\frac{\Theta_{sss}}2x^2+O_{\rm w}(\rho^3).
 \label{eq:local-fold-equation}
\end{align}
The generic cusp conditions
\begin{equation}
 \Theta_{s\lambda}\ne0,\qquad \Theta_{sss}\ne0
 \label{eq:generic-cusp}
\end{equation}
are strengthened to strict positivity by
Eq.~\eqref{eq:analytic-cusp-positivity} for every finite $\eps\geq0$.
Define
\begin{align}
 a_s&=\sqrt{\frac{2\lambda_c\Theta_{s\lambda}}{\Theta_{sss}}},
 & B&=\frac{\lambda_c\Theta_\lambda}{\Theta_c},\nonumber\\
 A_t&=\frac{2\lambda_c\Theta_{s\lambda}a_s}{3\Theta_c},
 & K&=\frac{\Theta_{sss}a_s^4}{24}
 =\frac{\lambda_c^2\Theta_{s\lambda}^2}{6\Theta_{sss}}
 =\frac{a_s\Theta_c A_t}{8},
 \label{eq:landau-coefficients}
\end{align}
and the normalized entropy displacement and conjugate field
\begin{equation}
 \varphi=\frac{x}{a_s}\in\mathbb C,\qquad
 H=\frac{2}{A_t}(\delta t_{\rm b}-B\delta p).
 \label{eq:complex-landau-parameter}
\end{equation}
Integration of Eq.~\eqref{eq:off-shell-stationarity} yields
\begin{equation}
 {
 \mathfrak g_{\rm off}=\mathfrak g_{\rm reg}
 +K\left(\frac{\varphi^4}{4}
 +\frac32\delta p\,\varphi^2-H\varphi\right)
 +O_{\rm w}(\rho^5).
 }
 \label{eq:landau-normal-form}
\end{equation}
The analytic background $\mathfrak g_{\rm reg}$ is independent of $x$.
The first omitted terms are displayed in
Eq.~\eqref{eq:app-landau-potential-series}; the normal form is a leading
weighted approximation, not an exact quartic representation of the enthalpy.

Its stationary cubic and discriminant are
\begin{align}
 0&=\varphi^3+3\delta p\,\varphi-H,
 \label{eq:landau-cubic}\\
 \Delta&=-27(H^2+4\delta p^3).
 \label{eq:landau-discriminant}
\end{align}
The three local sheets, with correlated Cardano branches, are derived in
Appendix~\ref{app:landau-derivation}; they do not enumerate all solutions
of the exact thermodynamic equation.  The fold condition $\Delta=0$ gives
\begin{align}
 \varphi_\pm&=\pm\sqrt{-\delta p},&
 H_\pm&=\mp2(-\delta p)^{3/2},& \delta p&<0,
 \label{eq:real-landau-folds}\\
 \varphi_\pm&=\pm i\sqrt{\delta p},&
 H_\pm&=\pm2i\,\delta p^{3/2},& \delta p&>0.
 \label{eq:complex-landau-folds}
\end{align}
Thus the real spinodals,
\begin{equation}
 x_\pm=\pm a_s\sqrt{-\delta p}+O(\delta p),
 \label{eq:real-fold-scaling}
\end{equation}
merge and continue as a complex conjugate pair. Higher-weight terms
deform the semicubical cusp in the original physical coordinates.

\begin{figure}[tbp]
 \centering
 \includegraphics[width=0.96\textwidth]{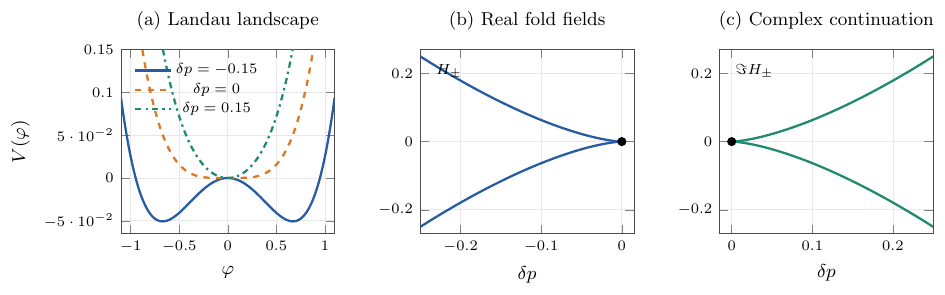}
 \caption{Universal Landau normal form.  (a) The dimensionless potential
 $V=\varphi^4/4+(3/2)\delta p\,\varphi^2-H\varphi$ at $H=0$ below, at, and
 above criticality.  (b) The real semicubical cusp
 $H_\pm=\mp2(-\delta p)^{3/2}$ for $\delta p<0$.  (c) Its supercritical
 continuation $\Im H_\pm=\pm2\delta p^{3/2}$ for $\delta p>0$; the two
 branches are complex conjugates.}
 \label{fig:landau-normal-form}
\end{figure}

The polynomial formulation also gives
\begin{equation}
 \frac{\Theta_{s\lambda}}{\Theta_{sss}}
 =\frac{\cF_\lambda}{\cF_{ss}}\bigg|_c.
 \label{eq:derivative-ratio}
\end{equation}
The analytic positivity proof and independently scaled Kerr endpoint imply
\begin{equation}
 K>0\quad(0\leq\eps<\infty),\qquad
 \bar K>0\quad\text{at the Kerr endpoint},
 \label{eq:no-tricritical-point}
\end{equation}
with $K_{\rm RN}=\sqrt6/48$ and $K|_{\eps=1}=0.0697151$.
Since the unnormalized quartic coefficient $\Theta_{sss}/96$ never
vanishes, a tricritical degeneration is excluded.  Figure~\ref{fig:landau-normal-form}
shows the potential and its real and complex fold fields.

\subsection{Phase structure implied by the Landau potential}
\label{sec:landau-phase-structure}

Curvatures and relative free energies distinguish the stationary roots as
stable, metastable, or unstable branches.  Their merger and equal depth
conditions determine the local spinodals and coexistence curve.

Write $q=\delta p$ and
$V_{\rm L}=\varphi^4/4+(3/2)q\varphi^2-H\varphi$.
For real $(\varphi,q,H)$, stability follows from
\begin{equation}
 V_{\rm L}''=3(\varphi^2+q),\qquad
 \partial_s^2\mathfrak g_{\rm off}
 =\frac{K}{a_s^2}V_{\rm L}''+O_{\rm w}(\rho^3).
 \label{eq:landau-stability-test}
\end{equation}
Inside the subcritical cusp, $q<0$ and $|H|<2(-q)^{3/2}$, the outer
roots are minima and the middle root is a maximum.  The lower minimum
is preferred, the higher minimum is metastable, and the maximum is
unstable.  On either fold the metastable minimum merges with the maximum;
outside the cusp only one real stationary minimum remains.  These are
stability statements for entropy fluctuations at fixed $(T_{\rm b},P,J,Q)$.

At $H=0$,
\begin{equation}
 \varphi=0,\qquad \varphi_{\rm eq}=\pm\sqrt{-3q},\qquad
 V_{\rm L}(\varphi_{\rm eq})=-\frac94q^2,\qquad q<0.
 \label{eq:landau-coexistence-minima}
\end{equation}
The equal depth minima represent coexisting small and large black holes
\cite{XuWuYang:2021,YangEtAl:2022}.  The field term $-H\varphi$ selects
the large branch for $H>0$ and the small branch for $H<0$, so crossing
$H=0$ produces a discontinuous switch.  Coexistence therefore lies inside
the spinodal cusp, rather than on its discriminant.

The leading quartic is symmetric, whereas the exact enthalpy is not.
Consequently $H=0$ fixes only the critical tangent:
\begin{equation}
 t_{\rm coex}=1+Bq+O(q^2),\qquad
 t_{\rm sp}^{\rm lo,hi}=1+Bq\mp A_t(-q)^{3/2}+O(q^2),\qquad q<0.
 \label{eq:landau-real-boundaries}
\end{equation}
The higher temperature spinodal terminates the metastable small branch;
the lower temperature one terminates the metastable large branch.
Exact coexistence obeys Eq.~\eqref{eq:coexistence-equations}, or equivalently
\begin{equation}
 0=\mathfrak g_{\rm off}(s_>)-\mathfrak g_{\rm off}(s_<)
 =\frac14\int_{s_<}^{s_>}
 [\Theta(s,\lambda;\eps)-\Theta_c t_{\rm coex}]\,\dd s.
 \label{eq:landau-exact-equal-area}
\end{equation}

Along coexistence, the leading entropy jump, latent heat, and barrier are
\begin{align}
 \Delta S=S_>-S_<&=2\sqrt3\,\pi Q^2a_s(-q)^{1/2}+O(|q|),
 \label{eq:landau-entropy-jump}\\
 L=T_{\rm coex}\Delta S
 &=2\sqrt3\,\pi Q^2T_ca_s(-q)^{1/2}+O(|q|),
 \label{eq:landau-latent-heat}\\
 \Delta\mathfrak g_{\rm barrier}
 &=\mathfrak g_{\rm off}(s_{\rm max})-\mathfrak g_{\rm off}(s_{\rm min})
 =\frac94Kq^2+O(|q|^{5/2}).
 \label{eq:landau-barrier}
\end{align}
Here $L$ is absorbed in the small-to-large transition, and the barrier
is in units of $Q$.  It does not fix a transition rate or hysteresis
trajectory without a dynamical prescription.  In terms of the reduced
latent heat of Eq.~\eqref{eq:latent-reduced},
\begin{equation}
 \mathcal L(p;\eps)
 =\frac{2\sqrt3\,a_s}{s_c}\sqrt{1-p}+O(1-p)
 =4\sqrt3\,A_z\sqrt{1-p}+O(1-p),\qquad p\to1^-.
 \label{eq:latent-critical}
\end{equation}
The edge amplitude is $A_z=a_s/(2s_c)$; the Kerr coefficient is
$2\sqrt3\,a_u/u_c$, with $a_u$ given in Section~\ref{sec:landau-family}.
Thus the exponent is common to the family, while its amplitude depends
on charge and rotation.

The stationary susceptibility is
\begin{equation}
 \chi_{\rm L}=\left(\frac{\partial\varphi}{\partial H}\right)_q
 =\frac{1}{3(\varphi^2+q)}.
 \label{eq:landau-susceptibility}
\end{equation}
At $H=0$ it equals $1/(3q)$ for $q>0$ and $1/(-6q)$ on the coexisting
branches for $q<0$.  Together with the jump $\Delta\varphi\propto(-q)^{1/2}$
and critical isobar $\varphi^3=H$, this gives the mean-field indices
\begin{equation}
 \beta_{\rm L}=\frac12,\qquad \gamma_{\rm L}=1,\qquad \delta_{\rm L}=3.
 \label{eq:landau-critical-indices}
\end{equation}
The endpoint is continuous.  For $q>0$, the real potential is strictly
convex: there is one minimum, no latent heat discontinuity, and a smooth
crossover.  Its complex folds are singularities of the continued equation
of state, not extra real phases.  These predictions are local; the exact
enthalpy determines the phase diagram away from criticality and the
separation between supercritical diagnostics.

\paragraph*{Landau coefficients across the charge--rotation family:}
\label{sec:landau-family}

\begin{figure}[tbp]
 \centering
 \includegraphics[width=0.94\textwidth]{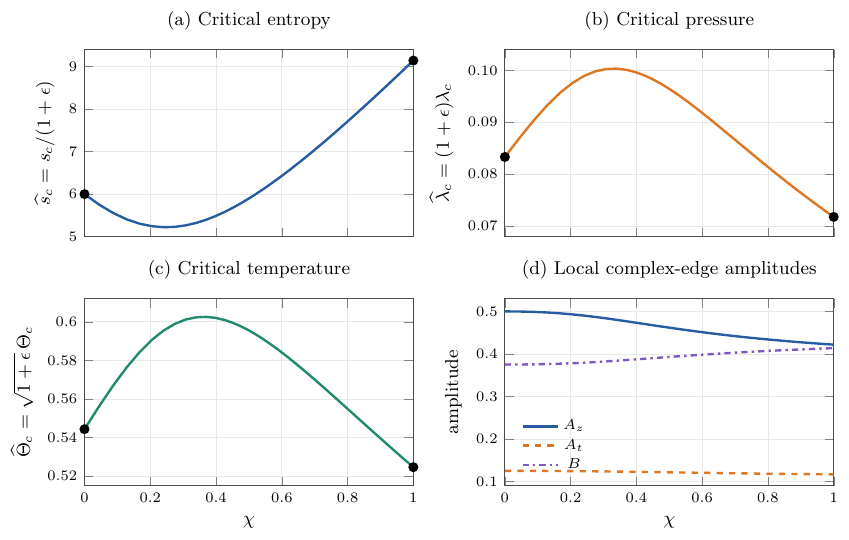}
 \caption{Charge--rotation critical family.  The compact coordinate
 $\chi=\eps/(1+\eps)$ maps RN--AdS to $\chi=0$ and Kerr--AdS to $\chi=1$.
 Panels (a)--(c) display finite scaled critical quantities; black
 points mark the analytic RN and independently computed Kerr endpoints.  Panel (d) shows the amplitudes
 defined in Eqs.~\eqref{eq:edge-amplitudes} and
 \eqref{eq:temperature-edge-scaling}.  Their smoothness and nonvanishing
 values support a generic cusp throughout the interpolation.}
 \label{fig:critical-family}
\end{figure}

The critical scales of Section~\ref{sec:critical-family} determine how
the cusp amplitudes interpolate between RN--AdS and Kerr--AdS.  In the
independently scaled rotating limit,
\begin{equation}
 \begin{aligned}
  \frac{a_s}{\eps}&\longrightarrow a_u=7.711686034,
  &\qquad
  A_z&\longrightarrow0.4217953882,\\
  A_t&\longrightarrow0.1166418360,
  &
  B&\longrightarrow0.4136372486,
 \end{aligned}
 \label{eq:kerr-landau-amplitudes}
\end{equation}
while $K/\sqrt\eps\to\bar K=0.05898884651$.  These rotating coefficients
are obtained from the Kerr enthalpy, rather than the RN--AdS specialization.
Figure~\ref{fig:critical-family} shows a smooth interpolation of the
critical scales and amplitudes in the compact variables of
Eq.~\eqref{eq:compact-critical-variables}; some critical scales vary
nonmonotonically.

Representative critical data and Landau amplitudes are collected in
Table~\ref{tab:critical-benchmarks}.

\begin{table}[tbp]
 \centering
 \small
 \begin{tabular}{c@{\hspace{0.8cm}}rrrrrr}
 \toprule
 $\eps$ & $s_c$ & $\lambda_c$ & $\Theta_c$ & $A_z$ & $A_t$ & $B$\\
 \midrule
 $0$   & 6.000000 & 0.0833333 & 0.5443311 & 0.500000 & 0.125000 & 0.375000\\
 $0.5$ & 7.963902 & 0.0668936 & 0.4916672 & 0.480876 & 0.123444 & 0.383710\\
 $1$   &11.795380 & 0.0482150 & 0.4210951 & 0.461748 & 0.121588 & 0.392885\\
 $2$   &20.490807 & 0.0294597 & 0.3319246 & 0.444800 & 0.119674 & 0.401419\\
 $10$  &93.252851 & 0.00690827& 0.1622772 & 0.426845 & 0.117352 & 0.410891\\
 \midrule
 Kerr limit & 9.141501 & 0.0717986 & 0.5246337 & 0.421795 & 0.116642 & 0.413637\\
 \bottomrule
 \end{tabular}
 \caption{Critical data and local complex edge amplitudes.  The last row
 uses the Kerr variables $(u_c,\bar\lambda_c,\bar\Theta_c)$ of
 Eq.~\eqref{eq:kerr-rescaling}.  The three amplitudes are the finite limits
 defined by Eqs.~\eqref{eq:edge-amplitudes},
 \eqref{eq:temperature-edge-scaling}, and
 \eqref{eq:kerr-landau-amplitudes}.}
 \label{tab:critical-benchmarks}
\end{table}

The finite, nonzero endpoint limits agree with the nondegeneracy proof.
Section~\ref{sec:complex-structure} uses these amplitudes to compare the
complex fold projection with the heat capacity ridge.

\section{Complex spinodals and the Widom surface}
\label{sec:complex-structure}

Above criticality, the physical spinodals continue as complex folds whose
real temperature projections define a Widom-type surface.  We derive
their near-critical scaling and compare the projected temperatures with
the heat capacity ridge.

\subsection{Complex continuation and the Widom surface}
\label{sec:complex-continuation}

We select the folds connected to the physical spinodals, construct their
real projection, and derive their near-critical scaling from the Landau cusp.

\subsubsection{Physical roots and the projected surface}
\label{sec:root-selection}

Root selection requires continuity from the two real spinodals and a
consistent choice of the square root branches in the thermodynamic functions.

\begin{figure}[tbp]
 \centering
 \includegraphics[width=0.91\textwidth]{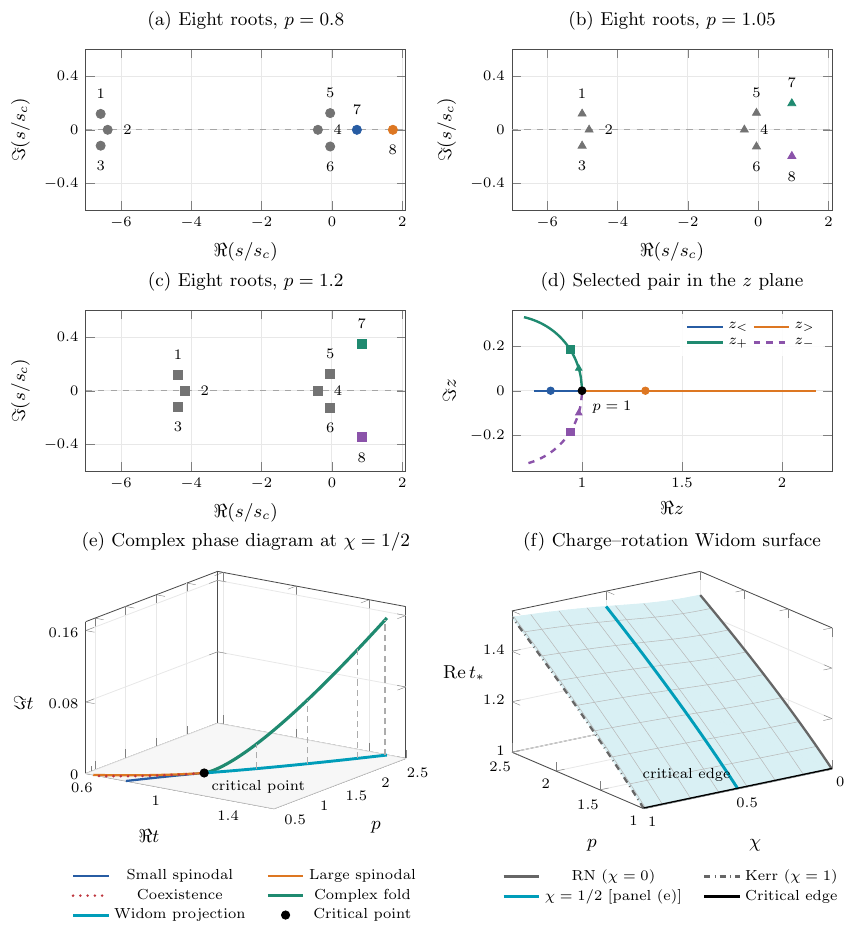}
 \caption{Complex fold geometry.  (a)--(c) All eight roots of
 $\cF(s,p\lambda_c;1)=0$ at $p=0.8,1.05,1.2$, respectively, on common
 axes.  Gray roots 1--6 have
 $\Re s<0$; colored roots 7 and 8 are, respectively, the small and large
 real spinodals in (a), and the upper and lower complex folds in (b) and (c).
 (d) The selected pair in $z=\sqrt{s/s_c}$: blue and orange denote
 $z_<,z_>$, while green and dashed purple denote $z_+,z_-$.
 In (d), colored circles mark $p=0.8$, triangles $p=1.05$, and squares
 $p=1.2$: these are the images of roots 7 and 8 in (a)--(c).
 The black dot at $z=1$ is their critical merger at $p=1$.
 (e) Complex diagram at $\eps=1$. Blue and orange retain the
 small and large spinodal colors; dotted red denotes coexistence, green the
 upper complex fold, cyan its Widom projection, and black the critical
 point.  Gray dashed segments join the fold to its projection on
 $\Im t=0$.  (f) The Widom-type surface $\Re t_*=t_W(p,\chi)$, with
 a mesh along the pressure and charge--rotation directions.
 RN and Kerr form the $\chi=0,1$ boundaries;
 $\chi=1/2$ is the slice in (e).  The black edge is $p=t_W=1$.
 Panel (e) shows the upper complex branch; both conjugate branches have
 the same real projection in (f).}
 \label{fig:complex-geometry}
\end{figure}

Keep $p$ and $\eps$ real and analytically continue the entropy variable in
Eq.~\eqref{eq:fold-equation}.  For $p<1$, select the two positive roots that
bound the unstable real branch.  At $p=1$ they meet at $s=s_c$.  Just above $p=1$, the nondegenerate cusp sends this pair off the real axis;
the reality of the coefficients makes the two continuations complex
conjugates.  Away from criticality we track this pair numerically and check
its residuals and square root sheets.  We denote the upper root and its reduced
temperature by
\begin{equation}
 s_*(p;\eps),
 \qquad
 \Im s_*>0,
 \qquad
 z_*=\sqrt{\frac{s_*}{s_c}},
 \qquad
 t_*=\frac{\Theta(s_*,p\lambda_c;\eps)}{\Theta_c}.
 \label{eq:complex-root-definitions}
\end{equation}
The square roots of both $s$ and $\mathcal D$ in $\Theta$, as well as
$z$, are chosen by continuity from their positive critical values.
The projected curve
\begin{equation}
 t_W(p;\eps)=\Re t_*(p;\eps),
 \qquad p>1,
 \label{eq:widom-definition}
\end{equation}
will be called a Widom-type line.  Varying $\eps$ turns this family of lines
into a surface in $(p,\chi,t_W)$.  More explicitly, the complex family and
its real projection are the two-dimensional sets
\begin{align}
 \Sigma_*&=\bigl\{(p,\chi,\Re t_*,\Im t_*):p>1,\ 0\leq\chi\leq1\bigr\},
 \notag\\
 \Sigma_W&=\bigl\{(p,\chi,t_W):p>1,\ 0\leq\chi\leq1\bigr\},
 \qquad \chi=\frac{\eps}{1+\eps}.
 \label{eq:widom-surface}
\end{align}
Here $\chi=1$ denotes the independently scaled Kerr endpoint, and the
numerical surface covers $1<p\leq2.5$.  Each fixed-$\chi$ curve projects
onto $\Im t=0$ as in Fig.~4 of Ref.~\cite{XuMann:2026}.
The full complex family has four real coordinates; Fig.~\ref{fig:complex-geometry}
therefore shows a fixed-ratio complex diagram and the projected surface
in $(p,\chi,\Re t_*)$.  Its critical edge $(p,\chi,t_W)=(1,\chi,1)$
represents different physical critical points, each reduced by its own
$P_c$ and $T_c$.

Root selection follows the physical spinodals, not proximity to the real
axis.  Figure~\ref{fig:complex-geometry}(a)--(c) resolves all eight roots of
Eq.~\eqref{eq:fold-equation} at $\eps=1$ for $p=0.8,1.05,1.2$ in separate
pressure snapshots.  These are poles of the heat capacity, not its zeros:
the numerator in Eq.~\eqref{eq:heat-capacity} is nonzero at every root
shown.  In each snapshot, the six gray roots have $\Re s<0$ and lie
outside the selected pair.
They describe neither positive entropy real black holes nor additional
physical critical points, which must satisfy
Eq.~\eqref{eq:critical-equations} on the physical branch.
Panels (d) and (e) show the selected roots and their temperature images;
panel (f) assembles their real projections into a surface.
The conjugate sheet has the opposite imaginary part and the same
projection.  This surface is a crossover diagnostic, not a coexistence
boundary; Section~\ref{sec:response-comparison} compares it with the
heat capacity ridge.

\begin{table}[tbp]
 \centering
 \small
 \setlength{\tabcolsep}{5pt}
 \renewcommand{\arraystretch}{1.15}
 \begin{tabular}{c r r r}
 \toprule
 Root & $p=0.8$ & $p=1.05$ & $p=1.2$\\
 \midrule
 1 & $-6.57679037+0.11893106\,i$ & $-5.00703026+0.11971508\,i$ & $-4.37917876+0.12019400\,i$ \\
 2 & $-6.37725841$ & $-4.80826966$ & $-4.18088927$ \\
 3 & $-6.57679037-0.11893106\,i$ & $-5.00703026-0.11971508\,i$ & $-4.37917876-0.12019400\,i$ \\
 4 & $-0.39888325$ & $-0.39234632$ & $-0.38864351$ \\
 5 & $-0.04922079+0.12481242\,i$ & $-0.04867011+0.12578582\,i$ & $-0.04832230+0.12637172\,i$ \\
 6 & $-0.04922079-0.12481242\,i$ & $-0.04867011-0.12578582\,i$ & $-0.04832230-0.12637172\,i$ \\
 \midrule
 7 & $0.71063037$ & $0.95752689+0.19709090\,i$ & $0.85109617+0.34710040\,i$ \\
 8 & $1.73401979$ & $0.95752689-0.19709090\,i$ & $0.85109617-0.34710040\,i$ \\
 \bottomrule
 \end{tabular}
 \caption{All eight roots $y=s/s_c$ of
 Eq.~\eqref{eq:fold-equation} at $\eps=1$ ($\chi=1/2$),
 rounded to eight decimal places. Root numbers match the labels within
 each snapshot in Fig.~\ref{fig:complex-geometry}(a)--(c).
 Roots 7 and 8 are the selected real spinodals at $p=0.8$ and the
 upper and lower complex folds at $p=1.05,1.2$; roots 1--6 have
 negative real entropy. At $p=1$, roots 7 and 8 coincide at $y=1$
 ($s=s_c$), forming the double root specified by
 Eq.~\eqref{eq:critical-equations} and marked by the black point in
 Fig.~\ref{fig:complex-geometry}(d).
 The unscaled roots are $s=s_cy$, with
 $s_c\simeq11.7953796613$.}
 \label{tab:figure4-roots}
\end{table}

\begin{figure}[tbp]
 \centering
 \includegraphics[width=\textwidth]{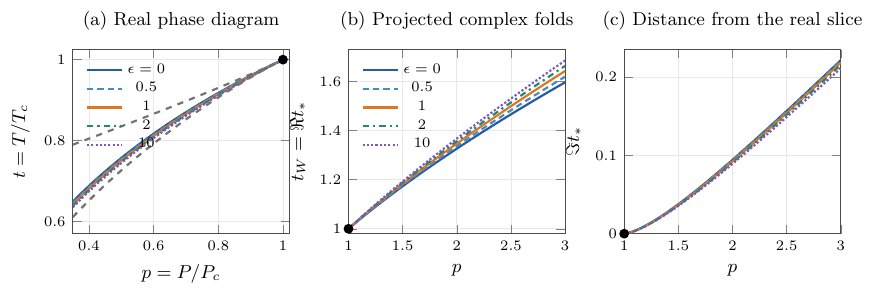}
 \caption{Real and complex phase structure for representative
 $\eps=0,0.5,1,2,10$.  (a) Real coexistence curves; the dashed gray curves
 are the two spinodals for $\eps=1$.  (b) Projection of the upper complex
 fold onto the real $(p,t)$ plane.  (c) Imaginary height of the same branch.
 The conjugate branch has the opposite imaginary part.}
 \label{fig:phase-structure}
\end{figure}

Figure~\ref{fig:phase-structure} compares coexistence, the projected
complex folds, and their imaginary heights across the charge--rotation
family.

For example, at $\eps=1$ and $p=1.2$ we find
\begin{equation}
 z_*=0.940810765+0.184468764\,i,
 \qquad
 t_*=1.076154456+0.010153578\,i.
 \label{eq:epsilon-one-example}
\end{equation}
This is the branch selected by continuation from the physical spinodals.

\subsubsection{Universal complex edge scaling}
\label{sec:edge-scaling}

The Landau cusp determines how far the selected folds move into the
complex entropy and temperature planes near criticality.  We extract the
scaling powers and the amplitudes that retain the charge--rotation
dependence.

The complex Landau fold~\eqref{eq:complex-landau-folds}, equivalently the
local equation~\eqref{eq:local-fold-equation}, immediately gives the upper
supercritical root
\begin{equation}
 x_*=i a_s\,\delta p^{1/2}+O(\delta p),
 \qquad \delta p>0.
 \label{eq:complex-entropy-scaling}
\end{equation}
Consequently,
\begin{equation}
 \Im z_*=A_z\,\delta p^{1/2}+O(\delta p^{3/2}),
 \qquad
 A_z=\frac{a_s}{2s_c}.
 \label{eq:edge-amplitudes}
\end{equation}
Substitution into Eq.~\eqref{eq:cusp-expansion}, together with the fold
condition, gives
\begin{equation}
 \Theta_{s\lambda}x_*\delta\lambda
 +\frac{1}{6}\Theta_{sss}x_*^3
 =\frac{2}{3}\Theta_{s\lambda}x_*\delta\lambda
 +O(\delta p^2).
 \label{eq:cubic-reduction}
\end{equation}
The reduced complex temperature therefore has the universal expansion
\begin{equation}
 {
 t_*=1+B\,\delta p+iA_t\,\delta p^{3/2}
 +O(\delta p^2),
 }
 \label{eq:temperature-edge-scaling}
\end{equation}
with
the same $B$ and $A_t$ that normalize the conjugate Landau field in
Eqs.~\eqref{eq:landau-coefficients} and
\eqref{eq:complex-landau-parameter}.
The lower branch has the opposite sign of the imaginary term.  The exponents
$1/2$ and $3/2$ follow only from the generic cusp conditions
\eqref{eq:generic-cusp}; charge and rotation enter through the amplitudes.

At the RN--AdS endpoint the fold equation can be solved exactly, yielding
\begin{equation}
 A_z=\frac12,
 \qquad
 A_t=\frac18,
 \qquad
 B=\frac38,
 \label{eq:rn-amplitudes}
\end{equation}
as shown in Appendix~\ref{app:rn}.  Numerical continuation gives the finite
$\eps$ values in Table~\ref{tab:critical-benchmarks}.  Logarithmic fits over
the near-critical interval give exponents $0.4995$ and $1.4997$ for every
representative $\eps$ shown in Fig.~\ref{fig:scaling-response}, consistent
with Eqs.~\eqref{eq:edge-amplitudes} and
\eqref{eq:temperature-edge-scaling}.

\subsection{Comparison with the heat capacity ridge}
\label{sec:response-comparison}

To compare the projected fold with a real thermodynamic diagnostic, we
follow the heat capacity maximum emerging from the critical divergence.
Their near-critical expansions determine both the common tangent and the
leading temperature separation.

\subsubsection{Exact response functions and ridge selection}
\label{sec:ridge-selection}

For $p>1$, the local maximum connected to criticality occurs at entropy
$s_{C_P}(p;\eps)$ and reduced temperature
\begin{equation}
 t_{C_P}(p;\eps)
 =\frac{\Theta(s_{C_P},p\lambda_c;\eps)}{\Theta_c}.
 \label{eq:cp-line}
\end{equation}
This is a conventional response function definition of a Widom line
\cite{OuyangYeLi:2024,DehyadegariSheykhi:2021,ZhaoEtAl:2025}.

The ridge is computed from the exact response functions at each pressure,
without a near-critical expansion.  Set $\lambda_p=p\lambda_c(\eps)$ and define
\begin{equation}
 \begin{aligned}
  \mathcal D_p(s)&=12\eps^2(\lambda_p s+3)
                  +(\lambda_p s^2+3s+3)^2,\\
  \mathcal N_p(s)&=\lambda_p^2s^4+4\lambda_p s^3
                  +(2\lambda_p+3)s^2-12\eps^2-3,\\
  \mathcal F_p(s)&=\cF(s,\lambda_p;\eps),
 \end{aligned}
 \label{eq:cp-exact-polynomials}
\end{equation}
where every coefficient of $\cF$ is given in
Eq.~\eqref{eq:fold-equation}.  The heat capacity and temperature are
\begin{equation}
 {\widehat C(s,p;\eps)
 =\frac{C_{P,J,Q}}{\pi Q^2}
 =\frac{2s\mathcal D_p(s)\mathcal N_p(s)}{\mathcal F_p(s)}},
 \qquad
 t(s,p;\eps)=\frac{\mathcal N_p(s)}
 {\Theta_c s^{3/2}\sqrt{\mathcal D_p(s)}}.
 \label{eq:cp-exact-functions}
\end{equation}
The positive square root is used on the real physical branch.
The dimensional values follow from
$S=\pi Q^2s$, $P=pP_c$, $J=\eps Q^2$,
$C_{P,J,Q}=\pi Q^2\widehat C$, and $T=T_ct$.
Normalization by the critical entropy instead gives
$C_{P,J,Q}/S_c=\widehat C/s_c$.

The maximum is located by differentiating the exact rational function.
With primes denoting $s$ derivatives at fixed $p,\eps$, define
\begin{equation}
 \begin{aligned}
 \mathcal R_p(s)
 &=[\mathcal D_p\mathcal N_p
   +s(\mathcal D_p'\mathcal N_p+\mathcal D_p\mathcal N_p')]
   \mathcal F_p
   -s\mathcal D_p\mathcal N_p\mathcal F_p',\\
 \partial_s\widehat C&=\frac{2\mathcal R_p}{\mathcal F_p^2},
 \qquad
 \left.\partial_s^2\widehat C\right|_{\mathcal R_p=0}
 =\frac{2\mathcal R_p'}{\mathcal F_p^2}.
 \end{aligned}
 \label{eq:cp-stationary-polynomial}
\end{equation}
Thus the numerical calculation follows the root $\mathcal R_p=0$ with
$s>0$, $\mathcal N_p>0$, $\mathcal F_p>0$, and $\mathcal R_p'<0$
that approaches $s_c$ as $p\to1^+$.  These conditions select a local maximum with positive temperature and positive heat capacity connected to criticality. Roots at $\mathcal F_p=0$ are excluded: they are the
spinodal divergences, not finite maxima.  Since $\partial_s t>0$ on
this stable branch, a maximum with respect to $s$ is also a maximum
with respect to $T$.  The stationary point equation involves only the
explicit polynomials in Eq.~\eqref{eq:cp-exact-polynomials} and their derivatives.

At the uncharged endpoint, the same construction uses $u$, $\bar\lambda$
and the polynomials $\mathcal D_K$, $\mathcal N_K$ in
Eqs.~\eqref{eq:D-kerr}~and~\eqref{eq:N-kerr}.  Explicitly,
\begin{equation}
 \frac{C_{P,J,0}}{\pi J}
 =\frac{2u\mathcal D_K\mathcal N_K}{\mathcal F_K},
 \qquad
 \mathcal F_K=2u\mathcal D_K\partial_u\mathcal N_K
 -\mathcal N_K(3\mathcal D_K+u\partial_u\mathcal D_K),
 \label{eq:cp-kerr-function}
\end{equation}
at $\bar\lambda=p\bar\lambda_c$; no division by $Q$ is required.

\subsubsection{Common tangent and quadratic temperature separation}
\label{sec:ridge-separation}

Put $q=p-1$ and write the reduced equation of state near criticality as
\begin{equation}
 t=1+Bq+aqx+\frac{b}{6}x^3+\frac{c}{2}q^2
   +\frac{d}{2}qx^2+\frac{e_4}{24}x^4+O_{\rm w}(\rho^5),
\end{equation}
where
\begin{equation}
 a=\frac{\lambda_c\Theta_{s\lambda}}{\Theta_c},\quad
 b=\frac{\Theta_{sss}}{\Theta_c},\quad
 c=\frac{\lambda_c^2\Theta_{\lambda\lambda}}{\Theta_c},\quad
 d=\frac{\lambda_c\Theta_{ss\lambda}}{\Theta_c},\quad
 e_4=\frac{\Theta_{ssss}}{\Theta_c}.
\end{equation}
The local heat capacity extremum obeys
\begin{equation}
 t_s^2-t\,t_{ss}
 =\frac{\mathcal R_p}{2\Theta_c^2s^5\mathcal D_p^3}=0.
 \label{eq:cp-exact-extremum}
\end{equation}
At criticality the derivative of its left-hand side with respect to $x$
equals $-b\ne0$.  The implicit function theorem therefore supplies a real
analytic extremum branch,
\begin{equation}
 x_{C_P}=-\frac{d}{b}q+O(q^2),\qquad
 t_{C_P}=1+Bq+\left(\frac c2-\frac{ad}{b}\right)q^2+O(q^3).
\end{equation}
It is a local maximum for sufficiently small $q>0$, since the second
entropy derivative of $C_{P,J,Q}/(\pi Q^2)$ has leading term $-b/(a^2q^2)<0$.
The upper complex fold instead has
\begin{equation}
 x_*=i\sqrt{\frac{2a}{b}}q^{1/2}
 +\left(-\frac db+\frac{e_4a}{3b^2}\right)q+O(q^{3/2}),
\end{equation}
which yields
\begin{equation}
 t_W=1+Bq+
 \left(\frac c2-\frac{ad}{b}+\frac{e_4a^2}{6b^2}\right)q^2+O(q^3).
\end{equation}
The real projection has integer powers because the two folds form a
conjugate Puiseux pair.  Their temperature difference is therefore
\begin{equation}
 {t_{C_P}-t_W=D_2q^2+O(q^3),\qquad
 D_2=-\frac{\lambda_c^2\Theta_{s\lambda}^2\Theta_{ssss}}
 {6\Theta_c\Theta_{sss}^2}>0.}
 \label{eq:quadratic-splitting}
\end{equation}
The strict sign follows from the squared enthalpy identities:
\begin{equation}
 \left.\frac{\Theta_{ssss}}{\Theta_{sss}}\right|_c
 =-\frac5{s_c}-\frac52\frac{h_{s,c}}{h_c}<0.
\end{equation}
For every member of the critical family, the heat capacity maximum lies
at a higher temperature than the complex projection sufficiently close
to criticality.  Equation~\eqref{eq:quadratic-splitting} gives both the
quadratic power and its positive coefficient.

\begin{figure}[tbp]
 \centering
 \includegraphics[width=\textwidth]{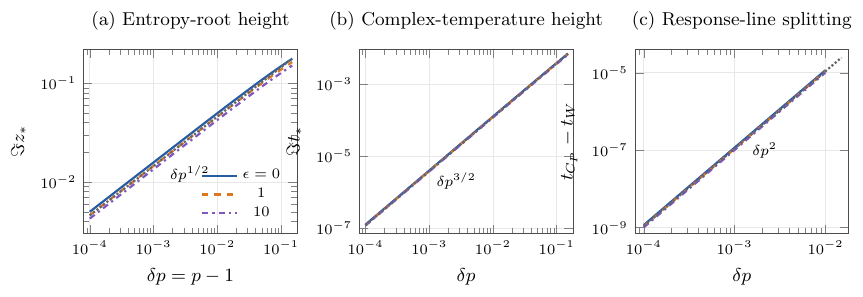}
 \caption{Universal near-critical scaling.  (a) The imaginary part of the entropy root follows $(p-1)^{1/2}$.  (b) The imaginary complex temperature height
 follows $(p-1)^{3/2}$.  (c) The temperature of the local $C_{P,J,Q}$ maximum
 and the projected complex fold have a common tangent and separate as
 $(p-1)^2$.  Dotted black lines indicate the stated powers.}
 \label{fig:scaling-response}
\end{figure}

For example, $D_2=15/128$ at $\eps=0$ and $D_2=0.1095876576$
at $\eps=1$.  Fits to 30 logarithmically spaced points over
$10^{-4}\leq q\leq10^{-2}$ give exponents between $2.000009$ and
$2.000119$ for $\eps=0,0.5,1,2,10$.  The ratio
$(t_{C_P}-t_W)/q^2$ agrees with the analytic coefficient to a relative
error below $7\times10^{-6}$ at $q=10^{-4}$.
Figure~\ref{fig:scaling-response}(c) displays this quadratic separation.

\subsubsection{Termination of the heat capacity ridge}
\label{sec:ridge-termination}

Equation~\eqref{eq:quadratic-splitting} is local.  Farther above criticality,
the heat capacity ridge can terminate when its local maximum merges with
a neighboring minimum~\cite{XuMann:2026}.  In terms of
Eq.~\eqref{eq:cp-stationary-polynomial}, the merger satisfies
\begin{equation}
 \mathcal R_{p_{\rm m}}(s_{\rm m})
 =\mathcal R'_{p_{\rm m}}(s_{\rm m})=0,
 \qquad s_{\rm m}>0,\quad
 \mathcal N_{p_{\rm m}}(s_{\rm m})>0,\quad
 \mathcal F_{p_{\rm m}}(s_{\rm m})>0.
 \label{eq:cp-ridge-termination}
\end{equation}
For RN--AdS, eliminating the entropy gives
$49p_{\rm m}^3+99p_{\rm m}^2-216p_{\rm m}-108=0$, with the physical
root $p_{\rm m}\simeq1.598858914$; at $\eps=1$, following the maximum
from criticality gives $p_{\rm m}\simeq1.636909891$.
The heat capacity remains finite and the equilibrium branch is smooth
at the merger, so this is the endpoint of a response diagnostic, not
another phase transition.  The complex fold projection continues beyond
these pressures in the numerical domain studied here.

 \subsection{Views at fixed angular momentum and fixed charge}
\label{sec:fixed-parameter-geometry}

The charge--rotation family can be viewed by holding either $J$ or $Q$
fixed across the plotted curves.  Common units retain the displacement
of the critical points and folds as the other conserved quantity varies.

\subsubsection{Coordinates at fixed angular momentum and fixed charge}
\label{sec:fixed-parameter-coordinates}

For fixed $J=J_0>0$, write $\eta=Q/\sqrt{J_0}$ and use
\begin{equation}
 u=\frac{S}{\pi J_0},\qquad
 \bar\lambda=8\pi PJ_0,\qquad
 \bar\Theta=4\pi\sqrt{J_0}\,T.
 \label{eq:fixed-J-coordinates}
\end{equation}
For $\eta>0$, the charge scaled fold solutions transform as
\begin{equation}
 \eps=\eta^{-2},\qquad
 u_*=\eta^2s_*,\qquad
 \bar\lambda=\frac{\lambda}{\eta^2},\qquad
 \bar\Theta_*=\frac{\Theta_*}{\eta}.
 \label{eq:fixed-J-fold-map}
\end{equation}
The endpoint $\eta=0$ is evaluated directly in the Kerr scaling variables.
For fixed $Q=Q_0>0$, the coordinates
\begin{equation}
 s=\frac{S}{\pi Q_0^2},\qquad
 \lambda=8\pi PQ_0^2,\qquad
 \Theta=4\pi Q_0T,\qquad \eps=\frac{J}{Q_0^2}
 \label{eq:fixed-Q-coordinates}
\end{equation}
already give the desired family, with RN--AdS at $\eps=0$.
In both views, the fold condition is
$(\partial T/\partial S)_{P,J,Q}=0$: both conserved quantities are held
fixed in the derivative, and one is varied only between curves.
Thus every curve belongs to the same canonical ensemble used above.

\subsubsection{Critical points and projected crossover surfaces}
\label{sec:fixed-parameter-surfaces}

\begin{figure}[tbp]
 \centering
 \includegraphics[width=\textwidth]{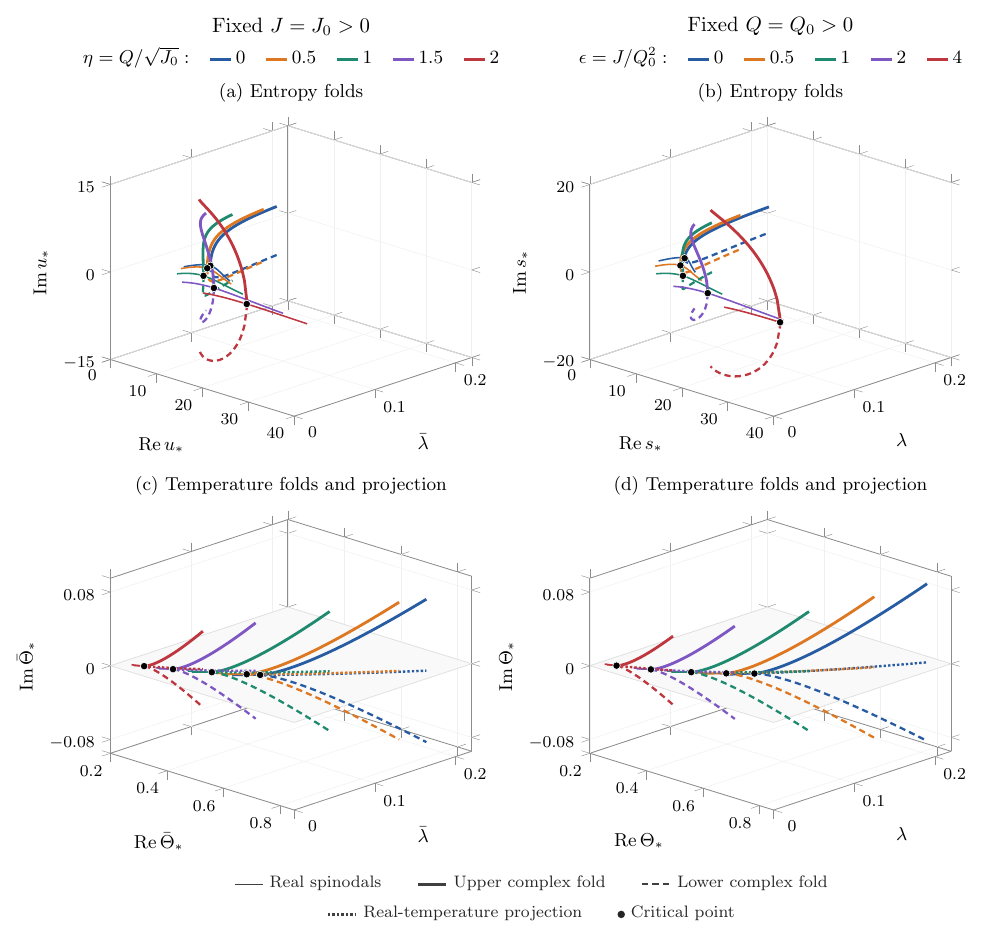}
 \caption{Selected folds at fixed $J=J_0$ (left) and fixed $Q=Q_0$
 (right) for $0.75\leq P/P_c\leq2.5$ on each curve.
 Panels (a) and (b) show the fold entropies; (c),(d) show their temperatures.
 Colors label the parameter values above each column; line styles
 distinguish the branches as indicated in the key.
 Black points are the unique critical points for the sampled $(J,Q)$
 pairs.  Dotted curves project the complex temperatures onto the pale
 $\mathrm{Im}\,T=0$ plane and define $T_W$.
 Coordinates follow Eqs.~\eqref{eq:fixed-J-coordinates} and
 \eqref{eq:fixed-Q-coordinates}. }
 \label{fig:fixed-parameter-geometry}
\end{figure}

Figure~\ref{fig:fixed-parameter-geometry} follows the two folds connected
to the physical critical point.  For $P<P_c$, they are real spinodals
that bound local stability in the entropy direction.  Coexistence is
instead fixed by equality of the Gibbs free energies.  At $P=P_c$ the
spinodals merge, and for $P>P_c$ they continue as conjugate complex
solutions.  These folds are branch points of the inverse equation of state $S(T)$,
while $T(S)$ remains regular there.  They do not represent additional
equilibrium black hole phases.

Each sampled $(J,Q)$ has one physical critical point
(Section~\ref{sec:analytic-critical-proof}), marked in black.  In the samples shown,
increasing $Q$ at fixed $J$, or $J$ at fixed $Q$, raises the critical
entropy and lowers the critical pressure and temperature.
Normalizing every curve by its own critical scales would hide this
motion by superposing the endpoints.  Colors identify parameters within
each column, rather than phases or a common $(J,Q)$ across columns;
the green curves coincide when $J_0=Q_0^2$ and $\eta=\eps=1$.

The conjugate temperatures define a common real projection,
\begin{equation*}
 T_*^\pm=T_R\pm iT_I,\qquad
 T_W(P;J,Q)=\Re T_*^\pm(P;J,Q).
\end{equation*}
The dotted curves define a crossover diagnostic, not real folds or
coexistence boundaries.  They share the heat capacity ridge's critical
value and tangent but differ at quadratic order in $p-1$
(Section~\ref{sec:ridge-separation}) and can persist beyond its endpoint.
Where the continuation exists, varying $Q$ at fixed $J$ gives a surface
in $(P,T,Q)$; varying $J$ at fixed $Q$ gives one in $(P,T,J)$.
Their near-critical boundary is the critical locus.
Figure~\ref{fig:fixed-parameter-geometry} shows five slices of each surface.
The rescaling~\eqref{eq:fixed-J-fold-map} preserves the cusp laws
$|\mathrm{Im}\,S_*|\propto(p-1)^{1/2}$ and
$|\mathrm{Im}\,T_*|\propto(p-1)^{3/2}$, with $p=P/P_c$ for each curve.
These laws hold only near criticality.

\FloatBarrier

\section{Discussion and conclusions}
\label{sec:discussion}

At fixed angular momentum and charge, the exact Kerr--Newman--AdS
enthalpy determines coexistence below criticality and the continuation of
the spinodals above it.  We have proved uniqueness and nondegeneracy of
the physical critical point, reduced coexistence to an exact quartic
equation, and derived the leading temperature difference between the
projected complex fold and the heat capacity ridge.

The critical quintic has one positive root for every finite
$\eps=J/Q^2\geq0$.  Both the quartic Landau coefficient and the pressure
coupling are strictly positive, excluding a tricritical degeneration
along this family; the independently scaled Kerr endpoint is likewise
nondegenerate.  Charge and rotation therefore change the scales and
amplitudes while preserving the local mean-field cusp.  The same
coefficients govern the square root vanishing of the latent heat below
criticality and the one-half and three-halves powers of the complex
entropy and temperature above it.

The common tangent construction extends the real analysis beyond the
Landau approximation.  Its quartic reduction determines both coexistence
entropies and $L=T_{\rm coex}\Delta S=\Delta M$ throughout $0<P<P_c$,
including the RN--AdS and Kerr--AdS limits, without a slow rotation
expansion.  Above criticality the real branch is smooth, with no entropy
jump or latent heat.  The complex roots instead locate singularities of
the analytically continued equation of state.

The projected fold and the local heat capacity ridge share a critical
tangent with the subcritical coexistence curve.  Their first difference
appears at quadratic order, as Eq.~\eqref{eq:quadratic-splitting} gives
\begin{equation*}
 t_{C_P}-t_W=D_2(\eps)(p-1)^2+O\!\left((p-1)^3\right),
 \qquad D_2(\eps)>0.
\end{equation*}
Thus the ridge lies at a higher temperature sufficiently close to
criticality; at the RN endpoint $D_2=15/128$.  Farther away, the response
maximum can disappear through the merger described by
Eq.~\eqref{eq:cp-ridge-termination}, while the complex projection continues.
The choice of diagnostic therefore affects both the crossover temperature
and its pressure range.  Varying the charge--rotation ratio turns the
projected lines into a surface.  Each point on its critical edge belongs
to a different fixed-$(J,Q)$ system; the surface is not a coexistence
boundary.

The proofs of uniqueness and nondegeneracy, the coexistence reduction,
and the local scaling laws are analytic.  Away from criticality, we have
followed the selected folds numerically over the stated pressure ranges.  Landau stability concerns entropy
variations at fixed bath temperature and $(P,J,Q)$, and does not establish
dynamical stability or transition rates.  Identifying the folds with
Lee--Yang or Fisher zeros would require a partition function and contour,
for example through a regulated gravitational saddle point integral.
Calculating the maxima of other response functions would test how their
crossover curves differ from the two studied here.  Changing the ensemble
would also change the allowed fluctuations and stability conditions.

\acknowledgments
\label{sec:acknowledgments}
BHL is supported by the National Research Foundation of Korea (NRF)
under grants RS-2020-NR049598, RS-2024-00339204, and RS-2026-25473640,
and by the Overseas Visiting Fellow Program of Shanghai University.
BHL thanks the Asia Pacific Center for Theoretical Physics (APCTP) and
the Korea Institute for Advanced Study (KIAS) for their hospitality during
visits in which part of this work was carried out.
HL is supported by the Basic Science Research Program through the
National Research Foundation of Korea (NRF), funded by the Ministry of
Education (NRF-2022R1I1A2063176), and by the Dongguk University Research
Fund of 2025.
ST thanks the Center for Quantum Spacetime (CQUeST), Sogang University,
for support for this work.

\textbf{Declaration of use of AI}: OpenAI ChatGPT was used for language
editing, \LaTeX{} consistency checks, and assistance with the initial
root-finding code for Fig.~4(a).  The code was checked manually.

\appendix

\section{Endpoint checks and phase structure benchmarks}
\label{app:rn}

Exact RN--AdS solutions check the real and complex folds, coexistence
curve, and entropy jump.  We derive coexistence from equal free energies
and the Maxwell construction, then compare RN--AdS and Kerr--AdS phase
diagrams at the same reduced pressures.

\paragraph*{Exact RN--AdS folds.}

The nonrotating limit makes the physical fold equation quadratic.  Its
closed form solution checks the root selection, complex scaling, and
near-critical temperature splitting of the general construction.

At $\eps=0$, let $A=\lambda s^2+3s+3$.  Then
\begin{equation}
 \mathcal D=A^2,\qquad
 \mathcal N=A(\lambda s^2+s-1),\qquad
 \cF=A^3(\lambda s^2-s+3).
\end{equation}
The six roots supplied by $A^3$ are removable factors in $\Theta_s$ and
are excluded before interpreting the polynomial roots as folds.  On the
RN sheet continued from positive entropy, the nonzero roots of $A$ are
also removable from the thermodynamic functions, not intrinsic branch
points.  The physical fold equation at this endpoint is the remaining
quadratic $\lambda s^2-s+3=0$.
Accordingly the temperature on the continued physical sheet is
\begin{equation}
 \Theta_{\mathrm{RN}}=\frac{\lambda s^2+s-1}{s^{3/2}}.
 \label{eq:rn-temperature}
\end{equation}
Using the critical values~\eqref{eq:rn-critical} and
$s=6z^2$, $\lambda=p/12$, one obtains
\begin{equation}
 t(z,p)=\frac{3pz^4+6z^2-1}{8z^3}.
 \label{eq:rn-reduced-temperature}
\end{equation}
The fold condition is exactly
\begin{equation}
 pz^4-2z^2+1=0,
 \label{eq:rn-fold}
\end{equation}
with
\begin{equation}
 z_{\pm}^2=\frac{1\pm\sqrt{1-p}}{p}.
 \label{eq:rn-fold-roots}
\end{equation}
For $p>1$, the upper branch has
\begin{equation}
 z_*^2=\frac{1+i\sqrt{p-1}}{p}.
 \label{eq:rn-complex-root}
\end{equation}
Expanding about $p=1+\delta p$ gives
\begin{align}
 z_*
 &=1+\frac{i}{2}\delta p^{1/2}-\frac{3}{8}\delta p
 +O(\delta p^{3/2}),
 \label{eq:rn-z-expansion}\\
 t_*
 &=1+\frac{3}{8}\delta p
 +\frac{i}{8}\delta p^{3/2}
 +O(\delta p^2),
 \label{eq:rn-t-expansion}
\end{align}
which proves Eq.~\eqref{eq:rn-amplitudes}.
The next real term gives an exact check of the response line splitting:
\begin{align}
 t_W&=1+\frac38\delta p-\frac9{128}\delta p^2+O(\delta p^3),\nonumber\\
 t_{C_P}&=1+\frac38\delta p+\frac3{64}\delta p^2+O(\delta p^3),\qquad
 D_2^{\rm RN}=\frac{15}{128}.
 \label{eq:rn-quadratic-check}
\end{align}

\paragraph*{Exact RN--AdS coexistence curve and entropies.}

At fixed charge, coexistence requires two distinct black holes with the
same temperature and Gibbs free energy.  We derive the resulting curve
using the exact off-shell potential and the Maxwell equal area
construction, and obtain both coexistence entropies.  In reduced
variables the pressure as a function of temperature is
\begin{equation}
 p_{\mathrm{coex}}(t)
 =\left[
 1-2\cos\left(
 \frac{\arccos(1-t^2)+\pi}{3}
 \right)
 \right]^2,
 \qquad 0<t<1.
 \label{eq:rn-coexistence}
\end{equation}

\paragraph*{Equal free energy construction.}
For RN--AdS, $z=\sqrt{S/S_c}=r_+/r_c$, with
$S_c=6\pi Q^2$ and $\cG_c=\sqrt6 Q/3$ for $Q>0$.
At a prescribed bath temperature $t_{\rm b}=T_{\rm b}/T_c$, the exact
off-shell potential is
\begin{equation}
 \widetilde g_{\rm off}(z;t_{\rm b},p)
 =\frac{M-T_{\rm b}S}{\cG_c}
 =\frac{pz^3+6z+z^{-1}}4-t_{\rm b}z^2.
 \label{eq:rn-offshell-exact}
\end{equation}
Stationarity gives
$\partial_z\widetilde g_{\rm off}=2z[t(z,p)-t_{\rm b}]=0$,
with $t(z,p)$ given in Eq.~\eqref{eq:rn-reduced-temperature}.
This potential retains the full entropy dependence, without a local
Landau expansion.

Let $z_<,z_>$ be the two positive coexistence radii, with common
temperature $t_0$ and reduced Gibbs energy $g_0$.  Equality of the two
potential values and stationarity make each radius a double zero of
$z(\widetilde g_{\rm off}-g_0)$.  The resulting quartic therefore
factorizes exactly:
\begin{equation}
 4z\bigl[\widetilde g_{\rm off}(z;t_0,p)-g_0\bigr]
 =pz^4-4t_0z^3+6z^2-4g_0z+1
 =p(z-z_<)^2(z-z_>)^2.
 \label{eq:rn-coexistence-factorization}
\end{equation}
By writing $U=z_<+z_>$ and $V=z_<z_>$, coefficient matching gives
\begin{equation}
 pV^2=1,\qquad 2pU=4t_0,\qquad
 p(U^2+2V)=6,\qquad 2pUV=4g_0.
 \label{eq:rn-coexistence-coefficients}
\end{equation}
Positivity of both radii selects $V=p^{-1/2}$.  Substituting
$U=2t_0/p$ into the relation for the quadratic coefficient yields
\begin{equation}
 {t_0^2=\frac{p(3-\sqrt p)}2},\qquad
 g_0=\sqrt{\frac{3-\sqrt p}{2}},\qquad 0<p<1.
 \label{eq:rn-coexistence-implicit}
\end{equation}
The positive square root gives the coexistence temperature.

To invert this relation, set $y=\sqrt p$.  Then
$y^3-3y^2+2t_0^2=0$.  The substitution $y=1-2\cos\theta$ gives
$\cos(3\theta)=t_0^2-1$.  The physical branch $0<y<1$ is
\begin{equation}
 \theta=\frac{\pi+\arccos(1-t_0^2)}3,
 \qquad \frac\pi3<\theta<\frac\pi2.
 \label{eq:rn-coexistence-cubic-branch}
\end{equation}
Squaring $y=1-2\cos\theta$ gives
Eq.~\eqref{eq:rn-coexistence}.  This branch approaches $p=0$ as
$t_0\to0$ and the critical point $p=t_0=1$ at its other endpoint.

\paragraph*{Maxwell construction in the temperature--entropy plane.}
The same coexistence condition follows from the first law at fixed
$P,Q$: $\dd M=T\,\dd S$.  Equality of $M-T_0S$ at the two endpoints is
therefore equivalent to
\begin{equation}
 T_0(S_>-S_<)=\int_{S_<}^{S_>}T(S,P,Q)\,\dd S.
 \label{eq:rn-maxwell-dimensional}
\end{equation}
Together with $T(S_<)=T(S_>)=T_0$, this is the equal area condition in
the temperature--entropy plane.  Since $S/S_c=z^2$, it becomes
\begin{align}
 t_0(z_>^2-z_<^2)
 &=\int_{z_<}^{z_>}2z\,t(z,p)\,\dd z\nonumber\\
 &=\frac14\left[
 p(z_>^3-z_<^3)+6(z_>-z_<)
 +\frac1{z_>}-\frac1{z_<}\right].
 \label{eq:rn-maxwell-integral}
\end{align}
The factor $2z$ is required by the entropy measure; an unweighted area
in the $(z,t)$ plane would give a different condition.  Dividing by
$z_>-z_<$ and using $U,V$ gives
\begin{equation}
 4t_0U=p(U^2-V)+6-\frac1V.
 \label{eq:rn-maxwell-sum-product}
\end{equation}
The equal endpoint temperatures, evaluated with
Eq.~\eqref{eq:rn-reduced-temperature}, imply
\begin{equation}
 U^2=6V^2+V-3pV^3,\qquad
 t_0=\frac U8\left(3p+\frac1{V^2}\right).
 \label{eq:rn-maxwell-endpoints}
\end{equation}
Substitution into Eq.~\eqref{eq:rn-maxwell-sum-product} reduces the
Maxwell condition to
\begin{equation}
 (pV^2-1)(pV^2-2V+1)=0.
 \label{eq:rn-maxwell-factorization}
\end{equation}
The second factor would give $U^2=4V$, or $z_<=z_>$, and is excluded for
two distinct coexisting phases.  Hence $V=p^{-1/2}$,
$U^2=6/p-2/\sqrt p$, and $t_0=pU/2$.  These reproduce
Eq.~\eqref{eq:rn-coexistence-implicit} and thus
Eq.~\eqref{eq:rn-coexistence}.

\begin{figure}[!t]
 \centering
 \includegraphics[width=\textwidth]{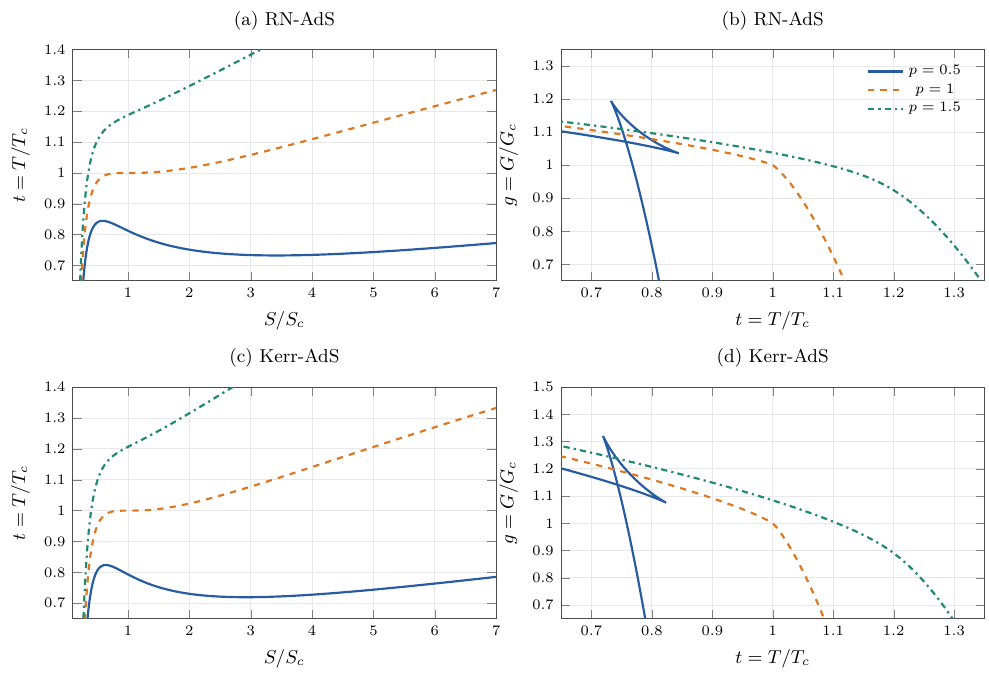}
 \caption{Endpoint phase structure benchmark from the exact enthalpy.
 Panels (a),(b) show RN--AdS ($J=0$, $Q>0$); panels (c),(d) show Kerr--AdS
 ($Q=0$, $J>0$), evaluated in its independent scaling variables.
 The same pressures $p=0.5,1,1.5$ are used in all panels.
 The horizontal entropy coordinate is $\widetilde S=S/S_c=z^2$,
 and each Gibbs curve is normalized by its endpoint's fixed critical
 value $\cG_c$.  The parametric Gibbs plots retain the unstable and
 metastable branches.  These reproduce the established endpoint
 thermodynamics~\cite{ChengWeiLiu:2016}; the complex continuation and
 response-line separation are analyzed in the main text.}
 \label{fig:endpoint-benchmark}
\end{figure}

\paragraph*{The two coexistence entropies.}
The sum and product determine the individual radii as the roots of
$z^2-Uz+V=0$.  Since
$U^2-4V=6(1-\sqrt p)/p>0$ for $0<p<1$, they are
\begin{equation}
 z_{\lessgtr}
 =\frac{\sqrt{6-2\sqrt p}\mp\sqrt{6-6\sqrt p}}{2\sqrt p}.
 \label{eq:rn-coexistence-radii}
\end{equation}
Squaring gives the reduced entropies explicitly:
\begin{equation}
 {\frac{S_{\lessgtr}}{S_c}
 =\frac{3-2\sqrt p\mp\sqrt{3(1-\sqrt p)(3-\sqrt p)}}p},
 \qquad S_c=6\pi Q^2.
 \label{eq:rn-coexistence-entropies}
\end{equation}
The minus sign gives the small black hole and the plus sign the large
black hole.  Both are stable minima: the factorization in
Eq.~\eqref{eq:rn-coexistence-factorization} implies
\begin{equation}
 \left.\partial_z^2\widetilde g_{\rm off}\right|_{z_{\lessgtr}}
 =\frac{p(z_>-z_<)^2}{2z_{\lessgtr}}>0.
 \label{eq:rn-coexistence-minima}
\end{equation}
These entropies describe phases with equal free energies and differ from the
spinodal entropies in Eq.~\eqref{eq:rn-fold-roots}.  They merge at
$S_<=S_>=S_c$ as $p\to1^-$.  Their difference is
\begin{equation}
 \frac{S_>-S_<}{S_c}
 =\frac{2\sqrt{3(1-\sqrt p)(3-\sqrt p)}}p.
 \label{eq:rn-coexistence-entropy-jump}
\end{equation}
Multiplication by $t_0$ gives
$L/(T_cS_c)=(3-\sqrt p)\sqrt{6(1-\sqrt p)/p}$,
recovering the RN--AdS latent heat in Eq.~\eqref{eq:latent-rn}.

\paragraph*{Direct comparison of the endpoint phase structures.}

The exact enthalpy also gives the isobars and Gibbs branches at both
endpoints.  The
comparison uses $\widetilde S=S/S_c$ and $\widetilde G=\cG/\cG_c$, with
$p=P/P_c$ and $t=T/T_c$ as in the main text.
For RN--AdS, Eqs.~\eqref{eq:enthalpy} and~\eqref{eq:gibbs} give
\begin{equation}
 t_{\rm RN}(\widetilde S,p)
 =\frac{3p\widetilde S^2+6\widetilde S-1}{8\widetilde S^{3/2}},
 \qquad
 \widetilde G_{\rm RN}(\widetilde S,p)
 =\frac{-p\widetilde S^2+6\widetilde S+3}{8\sqrt{\widetilde S}}.
 \label{eq:rn-endpoint-isobars}
\end{equation}

These expressions agree with the reduced equations in
Ref.~\cite{ChengWeiLiu:2016} and are independent of $Q$.
For Kerr--AdS we instead set $Q=0$ and use the independent $J$ scaling in
Eq.~\eqref{eq:kerr-rescaling}.  In terms of the squared enthalpy $h_K$
given below Eq.~\eqref{eq:kerr-critical-quintic}, define
$m_K=\sqrt{h_K}$ and $g_K=m_K-u\bar\Theta/4$.  Then
\begin{equation}
 t_K(\widetilde S,p)
 =\frac{\bar\Theta(u_c\widetilde S,p\bar\lambda_c)}{\bar\Theta_c},
 \qquad
 \widetilde G_K(\widetilde S,p)
 =\frac{g_K(u_c\widetilde S,p\bar\lambda_c)}
 {g_K(u_c,\bar\lambda_c)}.
 \label{eq:kerr-endpoint-isobars}
\end{equation}
The charge based variables are thus never evaluated at $Q=0$.

Figure~\ref{fig:endpoint-benchmark} shows both endpoints at the same
pressures $p=0.5,1,1.5$ in every panel.  It recovers the subcritical
oscillation and swallowtail, their merger at criticality, and the
monotone supercritical branches of Ref.~\cite{ChengWeiLiu:2016}.
For a direct numerical comparison with the displayed curves in that
reference, the pressure choices must be distinguished between panels.\footnote{
The displayed temperature--entropy curves in Figs.~1(a) and~3(a) of
arXiv:1603.08694v1 agree with $p=0.8,1,1.2$, whereas the
Gibbs--temperature curves in panels (b) agree with $p=0.5,1,1.5$, the
values stated in both captions.  Figure~\ref{fig:endpoint-benchmark}
uses the latter set consistently.}
The phase transition is determined by equality of $\cG$ on the two
locally stable branches, rather than by either spinodal temperature.
For example, at $p=0.5$ the conditions of equal temperature and equal free energy give
\begin{equation}
 \begin{aligned}
 (\widetilde S_<,\widetilde S_>,t_{\rm coex})_{\rm RN}
 &=(0.33275719,\,6.01038856,\,0.75711512),\\
 (\widetilde S_<,\widetilde S_>,t_{\rm coex})_K
 &=(0.40114968,\,4.98567024,\,0.74348161).
 \end{aligned}
 \label{eq:endpoint-coexistence-check}
\end{equation}
Both pairs lie on branches with positive heat capacity.  The RN
transition temperature also obeys the exact relation
$t_{\rm coex}=\sqrt{p(3-\sqrt p)/2}$, equivalent to
Eq.~\eqref{eq:rn-coexistence}.

These endpoint phase diagrams describe real small/large black hole
transitions.  Their complex continuation does not establish a
supercritical phase transition.

\FloatBarrier

\section{Derivation of the Landau normal form}
\label{app:landau-derivation}

We derive the weighted expansion, normalization, and local solutions
underlying Section~\ref{sec:landau}, retaining the first corrections to
the quartic potential.  The exact coexistence construction away from
criticality is given in Section~\ref{sec:latent-heat}.

\paragraph*{Exact potential and derivatives at the critical point.}

The first law fixes the entropy derivatives of the off-shell potential.
For $Q>0$, work at fixed $\eps=J/Q^2$ with the reduced squared enthalpy
\begin{equation}
 h(s,\lambda;\eps)
 =\frac{\lambda^2s^3}{36}+\frac{\lambda s^2}{6}
 +\frac{3+2\lambda}{12}s+\frac12+\frac{\lambda\eps^2}{3}
 +\frac{r^2}{4s},\qquad r^2=1+4\eps^2.
 \label{eq:app-landau-h}
\end{equation}
On the physical real branch, $\mathfrak m=\sqrt h>0$ and the first law gives
\begin{equation}
 \mathfrak m_s=\frac{h_s}{2\sqrt h}=\frac{\Theta}{4},
 \qquad \Theta=\frac{2h_s}{\sqrt h}.
 \label{eq:app-landau-first-law}
\end{equation}
The conditions $s_c,h_c>0$ select an analytic square root branch near
criticality, continued from the positive real mass.  The exact potential is
\begin{equation}
 \mathfrak g_{\rm off}(s;p,t_{\rm b},\eps)
 =\sqrt{h(s,p\lambda_c;\eps)}-\frac{\Theta_ct_{\rm b}}4s,
 \qquad t_{\rm b}=\frac{T_{\rm b}}{T_c}.
 \label{eq:app-landau-exact-potential}
\end{equation}
Entropy derivatives here hold $p,t_{\rm b},\eps$ fixed.  Consequently,
\begin{equation}
 \begin{aligned}
 4\partial_s\mathfrak g_{\rm off}&=\Theta-\Theta_ct_{\rm b},&
 4\partial_s^2\mathfrak g_{\rm off}&=\Theta_s,\\
 4\partial_s^3\mathfrak g_{\rm off}&=\Theta_{ss},&
 4\partial_s^4\mathfrak g_{\rm off}&=\Theta_{sss}.
 \end{aligned}
 \label{eq:app-landau-derivative-chain}
\end{equation}
At $(s,p,t_{\rm b})=(s_c,1,1)$, the first three derivatives vanish and
the fourth is positive, giving a quartic stationary point.  Changes in
the controls restore the linear and quadratic terms.

The required enthalpy derivatives are
\begin{equation}
 \begin{aligned}
 h_s&=\frac{\lambda^2s^2}{12}+\frac{\lambda s}{3}
       +\frac{3+2\lambda}{12}-\frac{r^2}{4s^2},&
 h_{ss}&=\frac{\lambda^2s}{6}+\frac\lambda3+\frac{r^2}{2s^3},\\
 h_{sss}&=\frac{\lambda^2}{6}-\frac{3r^2}{2s^4},&
 h_{ssss}&=\frac{6r^2}{s^5},\\
 h_\lambda&=\frac{\lambda s^3}{18}+\frac{s^2+s}{6}
                  +\frac{\eps^2}{3},&
 h_{s\lambda}&=\frac{\lambda s^2}{6}+\frac s3+\frac16,\\
 &&h_{ss\lambda}&=\frac{\lambda s+1}{3}.
 \end{aligned}
 \label{eq:app-landau-h-derivatives}
\end{equation}
For $\mathcal A=2hh_{ss}-h_s^2$, the identities
$\Theta_s=\mathcal A/h^{3/2}$ and $\mathcal A_s=2hh_{sss}$ reduce
criticality to $\mathcal A_c=h_{sss,c}=0$.  Evaluating derivatives at
$(s_c,\lambda_c;\eps)$ only after differentiation gives
\begin{equation}
 \begin{aligned}
 \Theta_{\lambda,c}
 &=\frac{2h_{s\lambda,c}}{\sqrt{h_c}}
   -\frac{h_{s,c}h_{\lambda,c}}{h_c^{3/2}},\\
 \Theta_{s\lambda,c}
 &=\frac{2h_{\lambda,c}h_{ss,c}
          +2h_ch_{ss\lambda,c}-2h_{s,c}h_{s\lambda,c}}{h_c^{3/2}},\\
 \Theta_{sss,c}&=\frac{2h_{ssss,c}}{\sqrt{h_c}}
                =\frac{12r^2}{s_c^5\sqrt{h_c}}>0.
 \end{aligned}
 \label{eq:app-landau-temperature-derivatives}
\end{equation}
Differentiating an expression already restricted to the critical curve
would instead give a derivative along that curve.

Using $\mathcal A_s|_c=0$ and
$\mathcal A(s,3r/s^2)=-f_r(s)/(16s^5)$ gives
\begin{equation}
 -\frac{2\lambda_c}{s_c}\mathcal A_{\lambda,c}
 =-\frac{f_r'(s_c)}{16s_c^5},\qquad
 \Theta_{s\lambda,c}
 =\frac{f_r'(s_c)}{32\lambda_cs_c^4h_c^{3/2}}>0.
 \label{eq:app-landau-pressure-positivity}
\end{equation}
The sign follows from the quintic proof in Section~\ref{sec:analytic-critical-proof}:
writing $f_r=s^5-c_3s^3-c_2s^2-c_1s-c_0$ gives
$s_cf_r'(s_c)=2c_3s_c^3+3c_2s_c^2+4c_1s_c+5c_0>0$.
Both the quartic term and the pressure coupling are therefore nonzero.

\paragraph*{Expansion and integration with cusp power counting.}

Balancing the temperature expansion determines the appropriate weights.
Put
\begin{equation}
 x=s-s_c,\qquad q=p-1,\qquad
 \tau=t_{\rm b}-1-Bq,\qquad
 B=\frac{\lambda_c\Theta_{\lambda,c}}{\Theta_c}.
 \label{eq:app-landau-controls}
\end{equation}
Removing the critical tangent leaves the balance $x^3\sim qx\sim\tau$,
hence
\begin{equation}
 x=O(\rho),\qquad q=O(\rho^2),\qquad \tau=O(\rho^3).
 \label{eq:app-landau-weights}
\end{equation}
A monomial $x^iq^j\tau^k$ has weight $i+2j+3k$; $O_{\rm w}(\rho^n)$
collects weights at least $n$.  Taylor expansion at fixed $\eps$ gives
\begin{equation}
 \begin{aligned}
 \Theta(s_c+x,\lambda_c+\lambda_cq;\eps)-\Theta_ct_{\rm b}
 ={}&-\Theta_c\tau+\lambda_c\Theta_{s\lambda,c}xq
       +\frac{\Theta_{sss,c}}6x^3\\
 &+\frac{\lambda_c^2\Theta_{\lambda\lambda,c}}2q^2
   +\frac{\lambda_c\Theta_{ss\lambda,c}}2x^2q
   +\frac{\Theta_{ssss,c}}{24}x^4
   +O_{\rm w}(\rho^5).
 \end{aligned}
 \label{eq:app-landau-temperature-series}
\end{equation}
Criticality removes the $x$ and $x^2$ terms, and $\tau$ absorbs the
term linear in $q$.  The two lines have weights three and four.

Choose the analytic background to be the exact value at $s=s_c$,
\begin{equation}
 \mathfrak g_{\rm reg}(q,\tau;\eps)
 =\mathfrak g_{\rm off}(s_c;1+q,1+Bq+\tau,\eps).
 \label{eq:app-landau-background}
\end{equation}
This $x$-independent term cancels from stationary conditions and free energy
differences.  Integrating Eq.~\eqref{eq:app-landau-derivative-chain} gives
\begin{equation}
 \mathfrak g_{\rm off}-\mathfrak g_{\rm reg}
 =\frac14\int_0^x
 \left[\Theta(s_c+\xi,\lambda_c+\lambda_cq;\eps)
       -\Theta_c(1+Bq+\tau)\right]\dd\xi.
 \label{eq:app-landau-integral}
\end{equation}
Integrating each Taylor monomial yields
\begin{equation}
 \begin{aligned}
 \mathfrak g_{\rm off}-\mathfrak g_{\rm reg}
 ={}&-\frac{\Theta_c}{4}\tau x
 +\frac{\lambda_c\Theta_{s\lambda,c}}8qx^2
 +\frac{\Theta_{sss,c}}{96}x^4\\
 &+\frac{\lambda_c^2\Theta_{\lambda\lambda,c}}8q^2x
 +\frac{\lambda_c\Theta_{ss\lambda,c}}{24}qx^3
 +\frac{\Theta_{ssss,c}}{480}x^5
 +O_{\rm w}(\rho^6).
 \end{aligned}
 \label{eq:app-landau-potential-series}
\end{equation}
Each coefficient includes the factor $1/4$ from the first law, its Taylor
factorial, and the integration factor.  Keeping only the first line
leaves $O_{\rm w}(\rho^5)$; in particular, $qx^3$ has weight five and
need not vanish.  No exact $x\mapsto-x$ symmetry is assumed.

\paragraph*{Normalization of the order parameter and field.}

Matching coefficients fixes the entropy, energy, and field scales.
Set $x=a_s\varphi$ and require
\begin{equation}
 \mathfrak g_{\rm off}-\mathfrak g_{\rm reg}
 =K\left(\frac{\varphi^4}{4}+\frac32q\varphi^2-H\varphi\right)
   +O_{\rm w}(\rho^5).
 \label{eq:app-landau-target}
\end{equation}
Matching the quartic, quadratic, and linear coefficients gives respectively
\begin{equation}
 \frac{\Theta_{sss,c}a_s^4}{96}=\frac K4,\qquad
 \frac{\lambda_c\Theta_{s\lambda,c}a_s^2}{8}=\frac{3K}{2},\qquad
 \frac{\Theta_ca_s\tau}{4}=KH.
 \label{eq:app-landau-matching}
\end{equation}
Their ratios determine
\begin{equation}
 \begin{aligned}
 a_s^2&=\frac{2\lambda_c\Theta_{s\lambda,c}}{\Theta_{sss,c}},&
 K&=\frac{\Theta_{sss,c}a_s^4}{24}
   =\frac{\lambda_c^2\Theta_{s\lambda,c}^2}{6\Theta_{sss,c}},\\
 A_t&=\frac{2\lambda_c\Theta_{s\lambda,c}a_s}{3\Theta_c},&
 H&=\frac{2\tau}{A_t},\qquad
 K=\frac{a_s\Theta_cA_t}{8}.
 \end{aligned}
 \label{eq:app-landau-scales}
\end{equation}
Choosing $a_s>0$ makes positive $\varphi$ a larger entropy displacement;
Eqs.~\eqref{eq:app-landau-temperature-derivatives} and
\eqref{eq:app-landau-pressure-positivity} give $A_t,K>0$.  The Jacobian
\begin{equation}
 \det\frac{\partial(q,H)}{\partial(p,t_{\rm b})}=\frac{2}{A_t}\ne0,
 \label{eq:app-landau-control-jacobian}
\end{equation}
and $\partial x/\partial\varphi=a_s\ne0$ establish nonsingular local
coordinates.  The quartic stationary point with independent linear and
quadratic controls is the $A_3$ cusp.  Equation~\eqref{eq:app-landau-potential-series}
retains the corrections that make the exact potential asymmetric.

\paragraph*{Stationary solutions, folds, and coexistence.}

We derive the cubic sheets and their mergers, then evaluate the
coexistence jump and barrier.  For the leading potential
$V_{\rm L}=\varphi^4/4+(3/2)q\varphi^2-H\varphi$,
\begin{equation}
 V_{\rm L}'=\varphi^3+3q\varphi-H,\qquad
 V_{\rm L}''=3(\varphi^2+q).
 \label{eq:app-landau-stationarity}
\end{equation}
Setting $\varphi=U+V$ gives
\begin{equation}
 \varphi^3+3q\varphi-H
 =U^3+V^3-H+3(UV+q)(U+V).
 \label{eq:app-landau-cardano-step}
\end{equation}
Thus choose $UV=-q$ and $U^3+V^3=H$.  Since $U^3V^3=-q^3$, the two
cubes are the roots of $z^2-Hz-q^3=0$, namely
\begin{equation}
 U^3=\frac H2+\sqrt{\frac{H^2}{4}+q^3},\qquad
 V^3=\frac H2-\sqrt{\frac{H^2}{4}+q^3}.
 \label{eq:app-landau-cardano-cubes}
\end{equation}
For $q\ne0$, choose one cube root $U$ and set $V=-q/U$; this enforces the
required correlation of branches.  The three solutions are
\begin{equation}
 \varphi_k=\omega^kU+\omega^{-k}V,\qquad
 \omega=e^{2\pi i/3},\qquad k=0,1,2.
 \label{eq:app-landau-cardano-roots}
\end{equation}
For $q=0$ they are the cube roots of $H$ and coincide at $H=0$.
Independent principal cube roots of the two expressions in
Eq.~\eqref{eq:app-landau-cardano-cubes} need not satisfy $UV=-q$;
Equation~\eqref{eq:app-landau-cardano-roots} describes only the local cubic sheets.

A fold is a simultaneous solution of $V_{\rm L}'=V_{\rm L}''=0$.
Eliminating $\varphi$ gives
\begin{equation}
 \varphi^2=-q,\qquad H=2q\varphi,
 \qquad H^2+4q^3=0.
 \label{eq:app-landau-fold-elimination}
\end{equation}
The discriminant $\Delta=-27(H^2+4q^3)$ therefore yields the real and
complex folds in Eqs.~\eqref{eq:real-landau-folds} and
\eqref{eq:complex-landau-folds}.
In entropy and temperature variables, these folds are
\begin{equation}
 \begin{aligned}
 s_\pm-s_c&=\pm ia_s q^{1/2}+O(q),&
 t_\pm&=1+Bq\pm iA_tq^{3/2}+O(q^2),\qquad q>0.
 \end{aligned}
 \label{eq:app-landau-physical-folds}
\end{equation}
Since $z=\sqrt{s/s_c}$, its expansion $z-1=x/(2s_c)+O(x^2)$ fixes
$A_z=a_s/(2s_c)$.  At $H=0$ and $q<0$, the cubic and curvature give
\begin{equation}
 \varphi_{\rm min}=\pm\sqrt{-3q},\qquad
 V_{\rm L}(\varphi_{\rm min})=-\frac94q^2,\qquad
 \varphi_{\rm max}=0,\quad V_{\rm L}(0)=0.
 \label{eq:app-landau-equal-minima}
\end{equation}
The separation of the two minima gives the entropy jump; the potential
difference between the maximum and either minimum gives the barrier:
\begin{equation}
 \begin{aligned}
 \Delta S&=2\sqrt3\,\pi Q^2a_s\sqrt{-q}+O(|q|),&
 \Delta\mathfrak g_{\rm barrier}&=\frac94Kq^2+O(|q|^{5/2}).
 \end{aligned}
 \label{eq:app-landau-jump-barrier}
\end{equation}
Multiplication by $T_{\rm coex}$ yields the latent heat.  The weight-five
corrections shift equal depths by $\tau=O(q^2)$, explaining why $H=0$
fixes only the coexistence tangent.  The exact condition is
Eq.~\eqref{eq:landau-exact-equal-area}.  The physical branch interpretation
is given in Section~\ref{sec:landau-phase-structure}; this thermodynamic
construction specifies neither an integration measure nor a gravitational
contour or transition dynamics.

\paragraph*{RN--AdS and Kerr--AdS endpoint normalizations.}

The endpoint coefficients check the derivative factors and normalization.
For RN--AdS, setting $\eps=0$ gives
\begin{equation}
 \begin{aligned}
 a_s&=6,\qquad B=\frac38,\qquad A_t=\frac18,\qquad
 K=\frac{\sqrt6}{48},\\
 \varphi&=\frac{s-6}{6},\qquad
 H=16\left(t_{\rm b}-1-\frac38q\right).
 \end{aligned}
 \label{eq:app-landau-rn-coefficients}
\end{equation}
These reproduce the RN--AdS potential expansion and give $A_z=1/2$.

For $Q=0$, use $u=S/(\pi J)$, $\bar\lambda=8\pi PJ$, and
$\bar\Theta=4\pi\sqrt J\,T$.  The independent normalization is
\begin{equation}
 \begin{aligned}
 h_K(u,\bar\lambda)=\frac{M^2}{J}
 &=\frac{\bar\lambda^2u^3}{36}+\frac{\bar\lambda u^2}{6}
   +\frac u4+\frac{\bar\lambda}{3}+\frac1u,\\
 \bar{\mathfrak g}_{\rm off}
 =\frac{M-T_{\rm b}S}{\sqrt J}
 &=\sqrt{h_K(u,p\bar\lambda_c)}-\frac{\bar\Theta_ct_{\rm b}}4u.
 \end{aligned}
 \label{eq:app-landau-kerr-potential}
\end{equation}
Repeating the steps above with
$(s,\lambda,\Theta)\mapsto(u,\bar\lambda,\bar\Theta)$ gives
\begin{equation}
 \begin{aligned}
 a_u^2&=\frac{2\bar\lambda_c\bar\Theta_{u\bar\lambda,c}}
                 {\bar\Theta_{uuu,c}},&
 \bar B&=\frac{\bar\lambda_c\bar\Theta_{\bar\lambda,c}}{\bar\Theta_c},\\
 \bar A_t&=\frac{2\bar\lambda_c\bar\Theta_{u\bar\lambda,c}a_u}
                     {3\bar\Theta_c},&
 \bar K&=\frac{\bar\Theta_{uuu,c}a_u^4}{24}.
 \end{aligned}
 \label{eq:app-landau-kerr-coefficients}
\end{equation}
With $\varphi_K=(u-u_c)/a_u$ and
$H_K=2(t_{\rm b}-1-\bar Bq)/\bar A_t$, the quartic is unchanged.
Here $\bar\Theta_{uuu,c}=48/(u_c^5\sqrt{h_{K,c}})>0$, and the Kerr
quintic~\eqref{eq:kerr-critical-quintic} gives
$\bar\Theta_{u\bar\lambda,c}>0$ by the argument of
Eq.~\eqref{eq:app-landau-pressure-positivity}.  The $Q=0$ endpoint
therefore has the same nondegenerate cusp in its own scaling variables.

\FloatBarrier

\bibliographystyle{JHEP}
\bibliography{knads4}

@article{Chamblin:1999,
  author        = {Chamblin, A. and Emparan, R. and Johnson, C. V. and Myers, R. C.},
  title         = {Charged {AdS} Black Holes and Catastrophic Holography},
  journal       = {Phys. Rev. D},
  volume        = {60},
  year          = {1999},
  pages         = {064018},
  eprint        = {hep-th/9902170},
  archivePrefix = {arXiv},
  doi           = {10.1103/PhysRevD.60.064018}
}

@article{KubiznakMann:2012,
  author        = {Kubiz{\v n}{\'a}k, D. and Mann, R. B.},
  title         = {$P$--$V$ Criticality of Charged {AdS} Black Holes},
  journal       = {JHEP},
  volume        = {07},
  year          = {2012},
  pages         = {033},
  eprint        = {1205.0559},
  archivePrefix = {arXiv},
  primaryClass  = {hep-th},
  doi           = {10.1007/JHEP07(2012)033}
}

@article{CveticEtAl:2011,
  author        = {Cveti{\v c}, M. and Gibbons, G. W. and Kubiz{\v n}{\'a}k, D. and Pope, C. N.},
  title         = {Black Hole Enthalpy and an Entropy Inequality for the Thermodynamic Volume},
  journal       = {Phys. Rev. D},
  volume        = {84},
  year          = {2011},
  pages         = {024037},
  eprint        = {1012.2888},
  archivePrefix = {arXiv},
  primaryClass  = {hep-th},
  doi           = {10.1103/PhysRevD.84.024037}
}

@article{Dolan:2011,
  author        = {Dolan, B. P.},
  title         = {Pressure and Volume in the First Law of Black Hole Thermodynamics},
  journal       = {Class. Quant. Grav.},
  volume        = {28},
  year          = {2011},
  pages         = {235017},
  eprint        = {1106.6260},
  archivePrefix = {arXiv},
  primaryClass  = {gr-qc},
  doi           = {10.1088/0264-9381/28/23/235017}
}

@article{CaldarelliEtAl:2000,
  author        = {Caldarelli, M. M. and Cognola, G. and Klemm, D.},
  title         = {Thermodynamics of Kerr--Newman--{AdS} Black Holes and Conformal Field Theories},
  journal       = {Class. Quant. Grav.},
  volume        = {17},
  year          = {2000},
  pages         = {399--420},
  eprint        = {hep-th/9908022},
  archivePrefix = {arXiv},
  doi           = {10.1088/0264-9381/17/2/310}
}

@article{ChengWeiLiu:2016,
  author        = {Cheng, P. and Wei, S.-W. and Liu, Y.-X.},
  title         = {Critical Phenomena in the Extended Phase Space of Kerr--Newman--{AdS} Black Holes},
  journal       = {Phys. Rev. D},
  volume        = {94},
  year          = {2016},
  pages         = {024025},
  eprint        = {1603.08694},
  archivePrefix = {arXiv},
  primaryClass  = {gr-qc},
  doi           = {10.1103/PhysRevD.94.024025}
}

@article{DehyadegariSheykhi:2021,
  author        = {Dehyadegari, A. and Sheykhi, A.},
  title         = {Thermodynamic Geometry and Phase Transition of Spinning {AdS} Black Holes},
  journal       = {Phys. Rev. D},
  volume        = {104},
  year          = {2021},
  pages         = {104066},
  eprint        = {2107.02915},
  archivePrefix = {arXiv},
  primaryClass  = {gr-qc},
  doi           = {10.1103/PhysRevD.104.104066}
}

@article{ZhaoEtAl:2025,
  author        = {Zhao, Z.-Q. and Nie, Z.-Y. and Zhang, J.-F. and Zhang, X.},
  title         = {Characterized behaviors of black hole thermodynamics in the supercritical region},
  journal       = {Chin. Phys. Lett.},
  volume        = {42},
  year          = {2025},
  pages         = {101102},
  eprint        = {2504.04995},
  archivePrefix = {arXiv},
  doi           = {10.1088/0256-307X/42/10/101102}
}

@article{OuyangYeLi:2024,
  author        = {Ouyang, X.-Y. and Ye, Q.-J. and Li, X.-Z.},
  title         = {Complex Phase Diagram and Supercritical Matter},
  journal       = {Phys. Rev. E},
  volume        = {109},
  year          = {2024},
  pages         = {024118},
  eprint        = {2303.16784},
  archivePrefix = {arXiv},
  doi           = {10.1103/PhysRevE.109.024118}
}

@article{XuMann:2026,
  author        = {Xu, Z.-M. and Mann, R. B.},
  title         = {Thermodynamic Supercriticality and Complex Phase Diagram for the {AdS} Black Hole},
  journal       = {Phys. Rev. Lett.},
  volume        = {136},
  year          = {2026},
  pages         = {041402},
  eprint        = {2504.05708},
  archivePrefix = {arXiv},
  doi           = {10.1103/c39y-zcz6}
}

@article{LiChenWuXu:2025,
  author        = {Li, Z.-Y. and Chen, X.-R. and Wu, B. and Xu, Z.-M.},
  title         = {Thermodynamic Supercriticality and Complex Phase Diagram for Charged Gauss--Bonnet {AdS} Black Holes},
  year          = {2025},
  eprint        = {2511.10357},
  archivePrefix = {arXiv},
  primaryClass  = {hep-th},
  url           = {https://arxiv.org/abs/2511.10357}
}

@article{OuZhangNie:2026,
  author        = {Ou, M. and Zhang, X. and Nie, Z.-Y.},
  title         = {Thermodynamic Supercriticality and Complex Phase Diagram for Charged {AdS} Black Holes in Trace Anomaly Gravity},
  year          = {2026},
  eprint        = {2608.09980},
  archivePrefix = {arXiv},
  primaryClass  = {gr-qc},
  url           = {https://arxiv.org/abs/2608.09980}
}

@article{XuWuYang:2021,
  author        = {Xu, Z.-M. and Wu, B. and Yang, W.-L.},
  title         = {van der Waals fluid and charged {AdS} black hole in the Landau theory},
  journal       = {Class. Quant. Grav.},
  volume        = {38},
  year          = {2021},
  pages         = {205008},
  eprint        = {2101.09456},
  archivePrefix = {arXiv},
  primaryClass  = {gr-qc},
  doi           = {10.1088/1361-6382/ac25dd}
}

@article{TsaiWuYang:2012,
  author        = {Tsai, Y.-D. and Wu, X. N. and Yang, Y.},
  title         = {Phase Structure of Kerr--{AdS} Black Hole},
  journal       = {Phys. Rev. D},
  volume        = {85},
  year          = {2012},
  pages         = {044005},
  eprint        = {1104.0502},
  archivePrefix = {arXiv},
  primaryClass  = {hep-th},
  doi           = {10.1103/PhysRevD.85.044005}
}

@article{YangEtAl:2022,
  author        = {Yang, S.-J. and Zhou, R. and Wei, S.-W. and Liu, Y.-X.},
  title         = {Kinetics of a phase transition for a Kerr--{AdS} black hole on the free-energy landscape},
  journal       = {Phys. Rev. D},
  volume        = {105},
  year          = {2022},
  pages         = {084030},
  eprint        = {2105.00491},
  archivePrefix = {arXiv},
  primaryClass  = {gr-qc},
  doi           = {10.1103/PhysRevD.105.084030}
}

@article{WeiChengLiu:2016,
  author        = {Wei, S.-W. and Cheng, P. and Liu, Y.-X.},
  title         = {Analytical and exact critical phenomena of $d$-dimensional singly spinning Kerr--{AdS} black holes},
  journal       = {Phys. Rev. D},
  volume        = {93},
  year          = {2016},
  pages         = {084015},
  eprint        = {1510.00085},
  archivePrefix = {arXiv},
  primaryClass  = {gr-qc},
  doi           = {10.1103/PhysRevD.93.084015}
}

@article{Thakur:KerrAdS5Complex,
  author        = {B.-H. Lee and H. Lee and S. Thakur},
  title         = {{Complex phase structure of Kerr--AdS$_5$ black holes: Critical saddles and Fisher zeros}},
  year          = {2026},
  eprint        = {2609.39085},
  archivePrefix = {arXiv},
  primaryClass  = {hep-th},
  url           = {https://arxiv.org/abs/2609.39085},
}

@article{KubiznakMannTeo:2017,
  author        = {D. Kubiz\v{n}\'ak and R. B. Mann and M. Teo},
  title         = {{Black hole chemistry: thermodynamics with Lambda}},
  journal       = {Class. Quant. Grav.},
  volume        = {34},
  year          = {2017},
  pages         = {063001},
  doi           = {10.1088/1361-6382/aa5c69},
  eprint        = {1608.06147},
  archivePrefix = {arXiv},
  primaryClass  = {hep-th},
  url           = {https://doi.org/10.1088/1361-6382/aa5c69},
}

@article{XuEtAl:2005Widom,
  author        = {L. Xu and P. Kumar and S. V. Buldyrev and S.-H. Chen and P. H. Poole and F. Sciortino and H. E. Stanley},
  title         = {{Relation between the Widom line and the dynamic crossover in systems with a liquid--liquid phase transition}},
  journal       = {Proc. Natl. Acad. Sci. U.S.A.},
  volume        = {102},
  number        = {46},
  year          = {2005},
  pages         = {16558--16562},
  doi           = {10.1073/pnas.0507870102},
  url           = {https://doi.org/10.1073/pnas.0507870102},
}

@article{Fisher:1978Edge,
  author        = {M. E. Fisher},
  title         = {{Yang--Lee Edge Singularity and $\phi^3$ Field Theory}},
  journal       = {Phys. Rev. Lett.},
  volume        = {40},
  number        = {25},
  year          = {1978},
  pages         = {1610--1613},
  doi           = {10.1103/PhysRevLett.40.1610},
  url           = {https://doi.org/10.1103/PhysRevLett.40.1610},
}

\end{document}